\documentclass[12pt,letterpaper]{article}
\usepackage{epsfig,rotating,setspace,latexsym,amsmath,epsf,amssymb,amsfonts,bm,theorem,cite,caption,subcaption,enumerate,longtable,accents}
\usepackage{algorithm,algorithmic,graphicx,epsf,authblk,epstopdf,url,color,multirow}
\usepackage{mathtools}
\usepackage{comment}
\usepackage{graphicx}
\usepackage{amsmath}
\usepackage{stackengine}
\usepackage{appendix}

\newtheorem{theorem}{Theorem}

\newtheorem{corollary}{Corollary}

\newtheorem{remark}{Remark}
\newtheorem{lemma}{Lemma}

\newenvironment{Proof}[1]{\medskip\par\noindent{\bf Proof:\,}\,#1}{{\mbox{\,$\blacksquare$}\par}}
\DeclareMathOperator{\diag}{diag}

\allowdisplaybreaks

\begin{document}

\title{Information-Theoretically Private Federated Submodel Learning with Storage Constrained Databases \thanks{This work was supported by ARO Grant W911NF2010142, and presented in part at IEEE ISIT 2022 and IEEE ISIT 2023.}}

\author{Sajani Vithana \qquad Sennur Ulukus\\
	\normalsize Department of Electrical and Computer Engineering\\
	\normalsize University of Maryland, College Park, MD 20742 \\
	\normalsize {\it spallego@umd.edu} \qquad {\it ulukus@umd.edu}}

\date{}
\maketitle

\vspace*{-1cm}

\begin{abstract}
In federated submodel learning (FSL), a machine learning model is divided into multiple submodels based on different types of data used for training. Each user involved in the training process only downloads and updates the submodel relevant to the user's local data, which significantly reduces the communication cost compared to classical federated learning (FL). However, the index of the submodel updated by the user and the values of the updates reveal information about the user's private data. In order to guarantee information-theoretic privacy in FSL, the model is stored at multiple non-colluding databases, and the user sends queries and updates to each database in such a way that no information is revealed on the updating submodel index or the values of the updates. In this work, we consider the practical scenario where the multiple non-colluding databases are allowed to have arbitrary storage constraints. The goal of this work is to develop read-write schemes and storage mechanisms for FSL that efficiently utilize the available storage in each database to store the submodel parameters in such a way that the total communication cost is minimized while guaranteeing information-theoretic privacy of the updating submodel index and the values of the updates. As the main result, we consider both heterogeneous and homogeneous storage constrained databases, and propose private read-write and storage schemes for the two cases.
\end{abstract}

\section{Introduction}

Federated learning (FL)\cite{FL1,FL2} is a widely used distributed learning framework, where millions of users collectively train a machine learning (ML) model by updating the model parameters using their private data in their local devices. Each user downloads the ML model, updates it, and uploads the updates back to a central server at multiple time instances in the training process. This results in a significantly large communication cost, which is one of the main drawbacks of FL. Many solutions have been proposed to reduce the communication overhead in FL, such as gradient sparsification \cite{sparse1,GGS,adaptive,conv,overtheair,timecorr}, where the users only download and upload a selected fraction of parameters and updates, gradient quantization \cite{qsl,fedpaq,qsgd,constraints} in which the updates are quantized to be represented using a fewer number of bits, and federated submodel learning (FSL) \cite{billion,secureFSL,paper1,dropout,ourICC,pruw_jpurnal,pruw,psufsl}, where the ML model is divided into multiple submodels based on different types of data used for training. In FSL, a given user only downloads and updates the submodel relevant to the user's local data, which significantly reduces the communication cost. FSL is especially useful in cases where the users participating in the training process have smaller amounts of data or if the data is class-specific.

One concern in FSL is the information leakage caused by specifying the index of the submodel updated by the user, as it directly reveals the kind of data that the user has. Moreover, it has been shown that the values of the updates uploaded by the users in distributed learning can be used to obtain information about the users' private data \cite{featureLeakage,InvertingGradients,DeepLeakage,BeyondClassRepresentatives,SecretSharer,comprehensive}. Privacy preserving FSL has been studied in the literature in terms of differential privacy \cite{secureFSL} as well as information-theoretic privacy \cite{paper1,dropout,ourICC,pruw_jpurnal}. The schemes presented in \cite{paper1,dropout,ourICC} guarantee information-theoretic privacy of the updating submodel index and the values of the updates sent by each individual user. Reference \cite{pruw_jpurnal} considers private FSL with sparsification which decreases the communication cost of the training process further.

Private FSL contains two phases, namely, the reading phase, in which the user downloads the required submodel without revealing its index, and the writing phase, in which the user uploads the updates back to the storage system without revealing the submodel index or the values of the updates. Note that the reading phase is identical to the problem of private information retrieval (PIR) \cite{PIR_ORI,PIR}. Private FSL is essentially an application of the basic problem in which a user downloads (reads) a required section of a data storage system, and uploads (writes) some content back to the selected section without revealing any information about the section index or the values of the content uploaded. We simply call this basic problem, private read-update-write (PRUW).

In this work, we consider the problem of PRUW with storage constrained databases, focusing on the application of private FSL. This problem is motivated by the fact that it requires multiple non-colluding databases to store the submodels to guarantee information-theoretic privacy of the user-required submodel index and the values of the updates. In practice, these non-colluding databases may have arbitrary storage constraints, which requires a flexible PRUW scheme that efficiently utilizes the available storage space in all databases to achieve the minimum possible communication cost in FSL. The main goal of this work is to determine storage mechanisms and compatible PRUW schemes that are applicable to any given set of arbitrary storage constraints. 

This work is closely related to the work presented in \cite{StorageConstrainedPIR, StorageConstrainedPIR_Wei, PIR_decentralized, efficient_storage_ITW2019, utah,coded,heteroPIR} in the PIR literature on storage constrained databases. In this work, we extend these ideas to PRUW with the goal of minimizing the total communication cost while guaranteeing the additional privacy and security requirements present in PRUW. Divided storage and coded storage are the two main approaches to storing data in databases with storage constraints. In divided storage, the data is divided into multiple segments, and each segment is only replicated in a subset of databases. In coded storage, multiple data points are combined into a single symbol using specific encoding structures, and stored at all databases. The concept of combining coded storage and divided storage to meet \emph{homogeneous} storage constraints in \emph{PIR} was introduced in \cite{efficient_storage_ITW2019}. It is shown that this combination results in better PIR rates compared to what is achieved by coded storage and divided storage individually. In this work, we explore such ideas for \emph{heterogeneous} storage in \emph{PRUW} in the context of a private FSL problem. In this paper, we propose a hybrid storage mechanism for private FSL that uses both divided and coded storage to store the submodel parameters, followed by a compatible PRUW scheme that achieves the minimum total communication cost within the algorithm for any given set of storage constraints, while guaranteeing the information-theoretic privacy of the user-required submodel index and the values of the updates.  

In this paper, we consider both heterogeneous and homogeneous storage constraints. The former corresponds to the case where different databases have different storage constraints, while the latter corresponds to the case with the same storage constraint across all databases. We provide two different schemes for the two settings, as they are rooted differently to increase the communication/storage efficiency of each case separately. Both schemes are composed of two main steps, namely, the storage mechanism, which assigns the content stored in each database, and the PRUW scheme, which performs the read-write process on the stored data. The PRUW scheme is based on the scheme presented in \cite{dropout, ourICC}, with the parameters optimized to achieve the minimum total communication cost, when the submodel parameters are $(K,R)$ MDS coded. In the heterogeneous case, the basic idea of the storage mechanism is to find the optimum $(K,R)$ MDS codes and the corresponding fractions of submodels stored using them. The general scheme that we propose in this work is applicable to any given set of storage constraints, and is based on a PRUW scheme that achieves lower and higher communication costs when the data is $(K,R)$ MDS coded with odd and even values of $R-K$, respectively. To this end, we show that the use of $(K,R)$ MDS codes with even $R-K$ can be avoided, and the total communication cost can be reduced if the given storage constraints satisfy a certain set of conditions. The class of homogeneous storage constraints satisfies these set of conditions. Hence, we discuss the case of homogeneous storage constraints separately in detail and propose a different storage mechanism that is more efficient compared to the general scheme designed for heterogeneous storage constraints.

\section{Problem Formulation}

We consider an FSL setting where the model to be trained consists of $M$ independent submodels, each containing $L$ parameters, taking values from a finite field $\mathbb{F}_q$. The submodels are stored in a system of $N$, $N\geq4$, non-colluding databases. Database $n$ has a storage capacity of $\mu(n) ML$ symbols, where $\mu(n)\in\left(0,1\right]$ for $n\in\{1,\dotsc,N\}$. In other words, each database must satisfy
\begin{align}
    H(S_n)\leq \mu(n) ML, \quad n\in\{1,\dotsc,N\},    
\end{align}
where $S_n$ is the content of database $n$. 

\begin{figure}[t]
    \centering
    \begin{subfigure}[b]{0.45\textwidth}
        \centering
        \includegraphics[width=\textwidth]{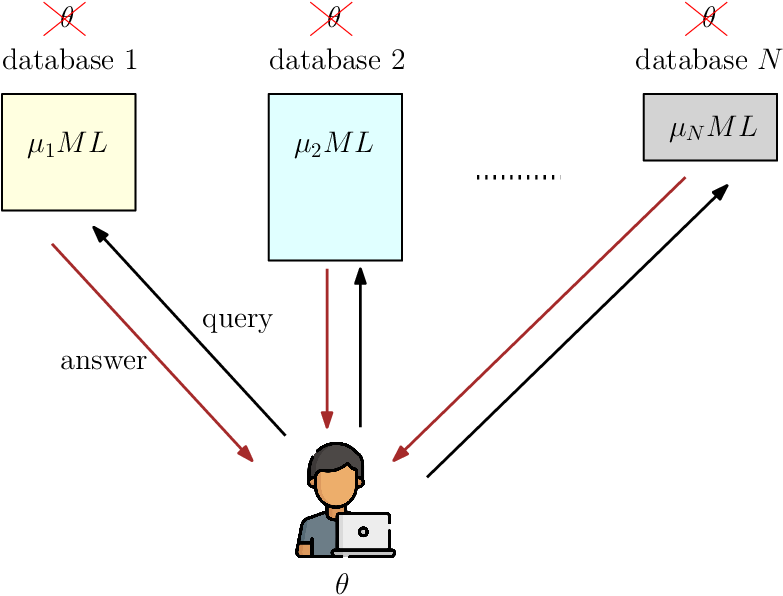}
        \caption{Reading phase.}
        \label{fig:read}
    \end{subfigure}
    \hfill
    \begin{subfigure}[b]{0.45\textwidth}
        \centering
        \includegraphics[width=\textwidth]{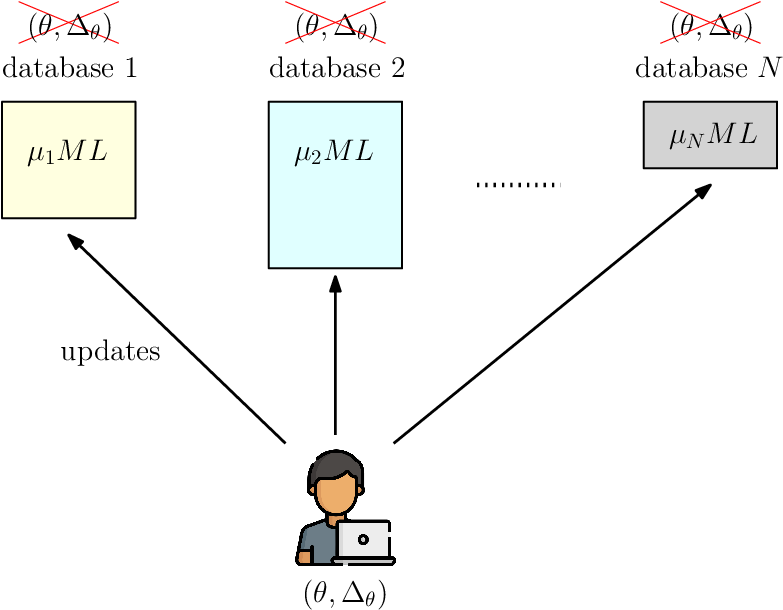}
        \caption{Writing phase.}
        \label{fig:write}
    \end{subfigure}
    \caption{A user reads a submodel, updates it, and writes it back to the databases.}
    \label{fig:model}
\end{figure}

The storage constraints $\mu(n)$ can be divided into two main categories, namely, heterogeneous constraints where there exists at least one pair of distinct storage constraints in the system, i.e., $\mu(\tilde{n})\neq\mu(n)$ for some $\tilde{n}\neq n$, $\tilde{n},n\in\{1,\dotsc,N\}$, and homogeneous constraints that satisfy $\mu(\tilde{n})=\mu(n)$ for all $\tilde{n},n\in\{1,\dotsc,N\}$. At any given time instance, a user downloads a required submodel by sending queries to all databases, without revealing the required submodel index to any of the databases. This is the reading phase of the PRUW process. The user then updates the submodel using the local data, and uploads the updates back to the same submodel in all databases without revealing the values of the updates or the updating submodel index to any of the databases. This is known as the writing phase. The two phases of the PRUW process are illustrated in Fig.~\ref{fig:model}. The following formal privacy and security constraints must be met in the PRUW process. 

\emph{Privacy of the submodel index:} At any given time $t$, no information on the indices of submodels updated up to time $t$, $\theta^{[1:t]}$, is allowed to leak to any of the databases,\footnote{The notation $[1:t]$ indicates all integers from $1$ to $t$.} i.e., for each $n$, $n\in\{1,\dotsc,N\}$,
\begin{align}
    I(\theta^{[1:t]};Q_n^{[1:t]},U_n^{[1:t]},S_n^{[0:t]})=0,
\end{align}
where $Q_n$ and $U_n$ are the queries and information on updates sent by the user to database $n$ at time instances denoted in square brackets in the reading and writing phases, respectively.

\emph{Privacy of the values of the updates:} At any given time $t$, no information on the values of the updates generated by the user up to time $t$, $\Delta_{\theta}^{[1:t]}$, is allowed to leak to any of the databases, i.e., for each $n$, $n\in\{1,\dotsc,N\}$,
\begin{align}
    I(\Delta_{\theta}^{[1:t]};U_n^{[1:t]},Q_n^{[1:t]},S_n^{[0:t]})=0.
\end{align}

\emph{Security of the submodels:} At any given time $t$, no information on the values of the parameters up to time $t$ in submodels is allowed to leak to any of the databases, i.e., for each $n$, $n\in\{1,\dotsc,N\}$,
\begin{align}
    I(W_{1:M}^{[0:t]};U_n^{[1:t]},Q_n^{[1:t]},S_n^{[0:t]})=0,
\end{align}
where $W_k$ represents the values of the parameters in the $k$th submodel at the time instance stated within brackets. Apart from the privacy and security guarantees, the process requires the following correctness conditions to be met in the reading and writing phases to ensure the reliability of the FSL process.

\emph{Correctness in the reading phase:} The user at time $t$ should be able to correctly download the required submodel from the answers received in the reading phase, i.e., 
\begin{align}
H(W_{\theta}^{[t-1]}|Q_{1:N}^{[t]},A_{1:N}^{[t]})=0,
\end{align}
where $A_n^{[t]}$ is the answer received from database $n$ at time $t$.

\emph{Correctness in the writing phase:} The submodels in all databases must be correctly updated as,
\begin{align}
    W_m^{[t]}=\begin{cases}
        W_m^{[t-1]}+\Delta_m^{[t]}, & \text{if 
 $m=\theta^{[t]}$}\\
        W_m^{[t-1]}, & \text{if $m\neq\theta^{[t]}$}.
    \end{cases}
\end{align}

The reading and writing costs are defined as $C_R=\frac{\mathcal{D}}{L}$ and $C_W=\frac{\mathcal{U}}{L}$, respectively, where $\mathcal{D}$ is the total number of symbols downloaded in the reading phase and $\mathcal{U}$ is the total number of symbols uploaded in the writing phase. The total cost $C_T$ is the sum of the reading and writing costs, i.e., $C_T=C_R+C_W$.

\section{Main Result}\label{main}

Theorems~\ref{thm1} and~\ref{thm0} in this section present the achievable total communication costs of the proposed schemes for heterogeneous and homogeneous storage constraints, respectively.

\begin{theorem}\label{thm1}
    In a PRUW setting with $N$ non-colluding databases having arbitrary heterogeneous storage constraints $\mu(n)$, $n\in\{1,\dotsc,N\}$, let $k=\frac{1}{\max_{n}\mu(n)}$, $p=\sum_{n=1}^N \mu(n)$, $r=kp$ and $s=\lfloor k\rfloor p$. Then, there exist a storage mechanism and a PRUW scheme that completely fills the $N$ databases and achieves a total communication cost of $\mathcal{C}=\min\{C_1,C_2\}$, where 
    \begin{align}\label{c1}
        C_1&=(\lceil s\rceil-s)C_T(\lfloor k\rfloor,\lfloor s\rfloor)+(s-\lfloor s\rfloor)C_T(\lfloor k\rfloor,\lceil s\rceil)\\
        C_2&=\alpha\beta C_T(\lfloor k\rfloor, \lfloor r\rfloor)+\alpha(1-\beta) C_T(\lfloor k\rfloor,\lceil r\rceil)+(1-\alpha)\delta C_T(\lceil k\rceil,\lfloor r\rfloor)\nonumber\\
        &\quad+(1-\alpha)(1-\delta) C_T(\lceil k\rceil,\lceil r\rceil)\label{c2}
    \end{align}
where $a$, $\beta$ and $\delta$ are given as, 
\begin{align}
    \alpha&=\begin{cases}\label{odd1}
        \frac{\lfloor k\rfloor(p\lceil k\rceil-\lceil r\rceil)}{\lceil k\rceil\lfloor r\rfloor-\lfloor k\rfloor\lceil r\rceil}, & \text{if $r-\lfloor r\rfloor>k-\lfloor k\rfloor$, $s\leq\lfloor r\rfloor$}\\
        \frac{\lfloor k\rfloor}{k}(\lceil k\rceil-k), & \text{else}
    \end{cases}\\
    \beta&=\begin{cases}
        \frac{\lceil r\rceil-r}{\lceil k\rceil-k}, & \text{if  $r-\lfloor r\rfloor>k-\lfloor k\rfloor$, $s>\lfloor r\rfloor$}\\
        1, & \text{else}
    \end{cases}\\
    \delta&=\begin{cases}\label{odd2}
        1-\frac{r-\lfloor r\rfloor}{k-\lfloor k\rfloor}, & \text{if  $r-\lfloor r\rfloor\leq k-\lfloor k\rfloor$}\\
        0, & \text{else}
    \end{cases},
\end{align}
 if $\lfloor r\rfloor-\lfloor k\rfloor$ is odd, and 
\begin{align}
    \alpha&=\begin{cases}\label{even1}
        \frac{\lfloor k\rfloor}{k}(\lceil k\rceil-k), & \text{if $r-\lfloor r\rfloor<\lceil k\rceil-k$}\\
        \frac{\lfloor k\rfloor(p\lceil k\rceil-\lfloor r\rfloor)}{\lceil k\rceil\lceil r\rceil-\lfloor k\rfloor\lfloor r\rfloor}, & \text{else}
    \end{cases}\\
    \beta&=\begin{cases}
        1-\frac{r-\lfloor r\rfloor}{\lceil k\rceil-k}, & \text{if  $r-\lfloor r\rfloor<\lceil k\rceil-k$}\\
        0, & \text{else}
    \end{cases}\\
    \delta&=1\label{even2}
\end{align}
if $\lfloor r\rfloor-\lfloor k\rfloor$ is even, with the function $C_T(a,b)$ defined as, 
\begin{align}\label{totalcost}
        C_T(a,b)=\begin{cases}
            \frac{4b}{b-a-1}, & \text{if $b-a$ is odd}\\
            \frac{4b-2}{b-a-2}, & \text{if $b-a$ is even}.
        \end{cases}
    \end{align}
\end{theorem}

\begin{remark}
    The intuition behind the value $p$ is the maximum number of times a given uncoded parameter can be replicated within the system of databases. The values $r$ and $s$ represent the number of times the parameters can be replicated if they are $k$ and $\lfloor k\rfloor$ coded, respectively. 
\end{remark}

\begin{figure}[t]
    \centering
    \includegraphics[scale=1]{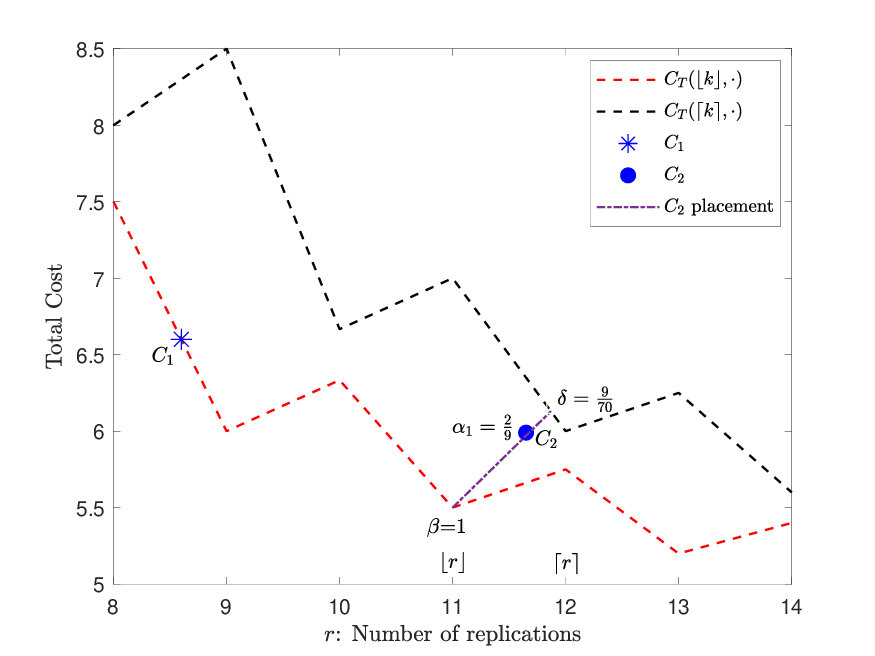}
    \caption{Example setting with $k=2.7$ and $p=4.3$.}
    \label{example}
    \vspace*{-0.5cm}
\end{figure}

\begin{remark}\label{rem1}
As an illustration of the main result, consider an example with $\max_n \mu(n)=0.37$ and $p=\sum_{n=1}^N \mu(n)=4.3$. Then, $k=2.7$, $r=11.61$ and $s=8.6$, which results in $C_1=6.6$ as shown by the blue star in Fig.~\ref{example}. Since $\lfloor r\rfloor-\lfloor k\rfloor=9$ is odd, $s=8.6<11=\lfloor r\rfloor$ and $r-\lfloor r\rfloor=0.61<0.7=k-\lfloor k\rfloor$, for the calculation of $C_2$, $\alpha$, $\beta$ and $\delta$ are given by $\alpha=\frac{\lfloor k\rfloor}{k}(\lceil k\rceil-k)=\frac{2}{9}$, $\beta=1$ and $\delta=1-\frac{r-\lfloor r\rfloor}{k-\lfloor k\rfloor}=\frac{9}{70}$, respectively. Hence, $C_2=5.99$ (blue dot in Fig.~\ref{example}). Note that $C_1$ is obtained by storing $\lceil s\rceil-s$ and $s-\lfloor s\rfloor$ fractions of parameters of all submodels using $(\lfloor k\rfloor, \lfloor s\rfloor)$ and $(\lfloor k\rfloor, \lceil s\rceil)$ MDS codes, respectively. Similarly, $C_2$ is obtained by storing $\alpha$, $(1-\alpha)\delta$ and $(1-\alpha)(1-\delta_1)$ fractions of all submodels using $(\lfloor k\rfloor,\lfloor r\rfloor)$, $(\lceil k\rceil,\lfloor r\rfloor)$ and $(\lceil k\rceil,\lceil r\rceil)$ MDS codes, respectively. This is illustrated in Fig.~\ref{example}. The minimum achievable cost for this example is $\min\{C_1,C_2\}=C_2$.
    
\end{remark}

\begin{remark}
    The values of $a$ and $b$ in the total cost expression in \eqref{totalcost} represent the coding parameter, i.e., the number of parameters linearly combined to a single symbol, and the number of databases each parameter is replicated at, respectively. The total communication cost decreases with the number of replications $b$, and increases with the coding parameter $a$, as shown in \eqref{totalcost}. However, when the number of replications is chosen as a linear function of the coding parameter, i.e., $r=kp$, $s=\lfloor k\rfloor p$, the total cost decreases with the coding parameter (see Section~\ref{encode}). Therefore, $C_2$ in the main result uses the maximum $k$ by considering both $\lfloor k\rfloor$ and $\lceil k\rceil$ when $k\notin\mathbb{Z}^+$. However, based on the fluctuating structure of the total cost in \eqref{totalcost} (dotted lines in Fig.~\ref{example}) for special cases of $k$ and $p$, a lower total cost can be achieved by only considering $\lfloor k\rfloor$. These cases are handled by $C_1$. 
\end{remark}

\begin{remark}
    For a given PRUW setting with arbitrary storage constraints $\{\mu(n)\}_{n=1}^N$, let the total available storage be $p=\sum_{n=1}^N \mu(n)$. Then, there exists an uncoded PRUW scheme (the proposed scheme in Section~\ref{general} with $k=1$ and $r=p$) that replicates $\lceil p\rceil-p$ and $p-\lfloor p\rfloor$ fractions of uncoded submodel parameters in $\lfloor p\rfloor$ and $\lceil p\rceil$ databases, respectively, achieving a total cost of $(\lceil p\rceil-p)C_T(1,\lfloor p\rfloor)+(p-\lfloor p\rfloor)C_T(1,\lceil p\rceil)$. Note that the total cost only depends on the sum of all arbitrary storage constraints (not on individual constraints), which is consistent with the corresponding result of private information retrieval with uncoded heterogeneous storage constraints. However, in PRUW, the total cost is minimized when $k\!=\!\frac{1}{\max_n\mu(n)}$ is maximized, i.e., when all storage constraints are equal. 
\end{remark}

\begin{theorem}\label{thm0}
Consider a PRUW setting consisting of $N$ non-colluding databases with a homogeneous storage constraint given by $\mu(n)=\mu\in\left[\frac{1}{N-3},1\right]$ for all $n\in\{1,\dotsc,N\}$. Then, there exists a scheme that satisfies all privacy and security constraints for any given $\mu\in\left[\frac{1}{N-3},1\right]$, and the resulting total communication cost is characterized by the lower convex hull of the following $(\bar{\mu},C_T(\bar{\mu}))$ pairs, where $\bar{\mu}$ and $C_T(\bar{\mu})$ are given by, 
\begin{align}
    \left(\tilde{\mu}=\frac{R}{NK_R},\quad C_T(\tilde{\mu})=\frac{4R}{R-K_R-1}\right),\quad R=4,\dotsc,N, 
    \quad K_R=1,\dotsc,R-3,
\end{align}
satisfying $(R-K_R-1)\!\!\mod2=0$.
\end{theorem}

\begin{remark}
    The scheme proposed for heterogeneous storage constraints is based on the idea of finding the optimum coding parameters in $(K,R)$ MDS coded storage that minimizes the total communication cost. However, as PRUW results in lower and higher communication costs for $(K,R)$ MDS codes with odd and even $R-K$, respectively, as shown in Fig.~\ref{example}, the costs presented in Theorem~\ref{thm1} are generally greater than what is presented in Theorem~\ref{thm0}, as the scheme proposed for homogeneous constraints are able to avoid the $(K,R)$ MDS codes with even $R-K$. However, a direct comparison cannot be made, as in general, the scheme proposed for heterogeneous constraints is not applicable for homogeneous constraints, as shown in the proofs of Lemmas~\ref{lem0} and~\ref{lem1} in Appendices~\ref{pflem1} and~\ref{pflem2}.  
\end{remark}

\section{Proposed Scheme: Heterogeneous Storage Constraints}

In this section, we present the proof of Theorem~\ref{thm1}. The proposed scheme for heterogeneous storage constrained PRUW consist of two components, namely, the storage mechanism and the PRUW scheme. The storage mechanism deals with finding the optimum coding parameters and optimum submodel partitions to be stored in each of the $N$ databases that result in the minimum reading and writing costs. The PRUW scheme is what states the steps of the read-write process based on the given storage structure. The PRUW scheme we use in this work is an optimized version of the general PRUW scheme in \cite{dropout}. In this section, we first present the optimized PRUW scheme, and then move on to the proposed storage mechanisms for any given set of storage constraints.

\subsection{General PRUW scheme}\label{pruw-general}

In this section, we use the general PRUW scheme proposed in \cite{dropout} for general $(K,R)$ MDS coded storage, and modify it by including the optimum subpacketization\footnote{Subpacketization is the number of parameters considered in each subpacket. A subpacket is a collection of parameters of all submodels, on which the scheme is defined. The scheme is then applied repeatedly on all subpackets identically.} and the optimum number of noise terms in storage to guarantee privacy, while minimizing the reading and writing costs.

\emph{Storage:} The contents of a single subpacket in database $n$, $n\in\{1,\dotsc,R\}$, with a subpacketization of $y$ and $x+1$ noise terms is given by,\footnote{Here, we consider storing each subpacket in only $R$ databases out of $N$, since all subpackets cannot be stored in all databases due to limited storage capacities in databases. $R$ is a variable which will be optimized later.}
\begin{align}
S_n=\begin{bmatrix}
    \sum_{i=1}^K\frac{1}{f_{1,i}-\alpha_n}\begin{bmatrix}
    W_{1,1}^{[i]}\\\vdots\\W_{M,1}^{[i]}
    \end{bmatrix}+\sum_{j=0}^x\alpha_n^j Z_{1,j}\\
    \vdots\\
    \sum_{i=1}^K\frac{1}{f_{y,i}-\alpha_n}\begin{bmatrix}
    W_{1,y}^{[i]}\\\vdots\\W_{M,y}^{[i]}
    \end{bmatrix}+\sum_{j=0}^x\alpha_n^j Z_{y,j}
    \end{bmatrix},\label{storage}
\end{align}
where $W_{m,j}^{[i]}$ is the $i$th parameter of the $j$th coded symbol of submodel $m$ (in the subpacket considered), $Z_{i,j}$ are random noise vectors of size $M\times1$ and $f_{i,j}$s and $\alpha_n$s are globally known distinct constants from $\mathbb{F}_q$. Note that the $j$th coded data symbol (without the noise) is stored in terms of a $(K,R)$ MDS code, with the generator matrix given by,
\begin{align}
    G=\begin{bmatrix}
        \frac{1}{f_{j,1}-\alpha_1} & \frac{1}{f_{j,2}-\alpha_1} & \dotsc & \frac{1}{f_{j,K}-\alpha_1}\\
        \vdots & \vdots & \vdots & \vdots\\ 
        \frac{1}{f_{j,1}-\alpha_R} & \frac{1}{f_{j,2}-\alpha_R} & \dotsc & \frac{1}{f_{j,K}-\alpha_R}
    \end{bmatrix}, \quad j\in\{1,\dotsc,y\}.
\end{align}
The variables $K$, $R$, $x$ and $y$ are optimized to achieve minimum reading and writing costs at the end of this section.

\emph{Reading phase:} In the reading phase, the user sends queries to all databases to download the required submodel $W_{\theta}$. The queries sent to database $n$, $n\in\{1,\dotsc,R\}$, to download the $K\times y$ parameters of each subpacket of the required submodel is given by,
\begin{align}           
    Q_{n,\ell}=\begin{bmatrix}
    \frac{\prod_{i=1,i\neq \ell}^{K}(f_{1,i}-\alpha_n)}{\prod_{i=1,i\neq \ell}^{K}(f_{1,i}-f_{1,\ell})}e_M(\theta)+\prod_{i=1}^K (f_{1,i}-\alpha_n)\tilde{Z}_{1,\ell}\\
    \vdots\\
    \frac{\prod_{i=1,i\neq \ell}^{K}(f_{y,i}-\alpha_n)}{\prod_{i=1,i\neq \ell}^{K}(f_{y,i}-f_{y,\ell})}e_M(\theta)+\prod_{i=1}^K (f_{y,i}-\alpha_n)\Tilde{Z}_{y,\ell}
\end{bmatrix}, \quad \ell\in\{1,\dotsc,K\},\label{query}
\end{align}
where $e_M(\theta)$ is the all zeros vector of size $M\times1$ with a 1 at the $\theta$th position and $\Tilde{Z}_{j,\ell}$ are random noise vectors of size $M\times1$. Then, database $n$, $n\in\{1,\dotsc,R\}$, sends the corresponding answers given by,
\begin{align}\label{ans}
    A_{n,\ell}&=S_n^T Q_{n,\ell}, \quad \ell\in\{1,\dotsc,K\}\\
    &=\begin{bmatrix}
    \sum_{i=1}^K\frac{1}{f_{1,i}-\alpha_n}\begin{bmatrix}
    W_{1,1}^{[i]}\\\vdots\\W_{M,1}^{[i]}
    \end{bmatrix}+\sum_{j=0}^x\alpha_n^j Z_{1,j}\\
    \vdots\\
    \sum_{i=1}^K\frac{1}{f_{y,i}-\alpha_n}\begin{bmatrix}
    W_{1,y}^{[i]}\\\vdots\\W_{M,y}^{[i]}
    \end{bmatrix}+\sum_{j=0}^x\alpha_n^j Z_{y,j}
    \end{bmatrix}^T\begin{bmatrix}
    \frac{\prod_{i=1,i\neq \ell}^{K}(f_{1,i}-\alpha_n)}{\prod_{i=1,i\neq \ell}^{K}(f_{1,i}-f_{1,\ell})}e_M(\theta)+\prod_{i=1}^K (f_{1,i}-\alpha_n)\tilde{Z}_{1,\ell}\\
    \vdots\\
    \frac{\prod_{i=1,i\neq \ell}^{K}(f_{y,i}-\alpha_n)}{\prod_{i=1,i\neq \ell}^{K}(f_{y,i}-f_{y,\ell})}e_M(\theta)+\prod_{i=1}^K (f_{y,i}-\alpha_n)\Tilde{Z}_{y,\ell}
\end{bmatrix}\\
    &=\sum_{j=1}^y\frac{1}{f_{j,\ell}-\alpha_n}W_{\theta,j}^{[\ell]}+P_{\alpha_n}(K+x), \quad \ell\in\{1,\dotsc,K\}, \label{last_ans}
\end{align}
where $P_{\alpha_n}(h)$ is a polynomial in $\alpha_n$ of degree $h$, and \eqref{last_ans} is a result of
\begin{align}
    \frac{1}{f_{j,i}-\alpha_n}\times\frac{\prod_{i=1,i\neq \ell}^{K}(f_{j,i}-\alpha_n)}{\prod_{i=1,i\neq \ell}^{K}(f_{j,i}-f_{j,\ell})}=\begin{cases}
        \frac{1}{f_{j,\ell}-\alpha_n}+P_{\alpha_n}(K-2), & \text{ if $i=\ell$}\\
        P_{\alpha_n}(K-2), & \text{if $i\neq\ell$},
    \end{cases}
\end{align}
which is obtained by applying \cite[Lem.~1]{pruw_jpurnal}. Note that the terms corresponding to $W_{\theta,j}^{[i\neq \ell]}$ for $j=1,\dotsc,y$ and $i=1,\dotsc,K$ in the calculation of the answers in \eqref{last_ans} are included in the combined noise polynomial $P_{\alpha_n}(K+x)$. Using the $K$ answers from each database, the user obtains the $K\times y$ parameters of each subpacket of the required submodel $W_{\theta}$ by solving,
\begin{align}\label{mat}
    \begin{bmatrix}
        A_{1,\ell}\\\vdots\\A_{R,\ell}
    \end{bmatrix}=\begin{bmatrix}
        \frac{1}{f_{1,\ell}-\alpha_1} & \dotsc & \frac{1}{f_{y,\ell}-\alpha_1} & 1 & \alpha_1 & \dotsc & \alpha_1^{K+x}\\
        \vdots & \vdots &\vdots &\vdots &\vdots &\vdots &\vdots\\
        \frac{1}{f_{1,\ell}-\alpha_R} & \dotsc & \frac{1}{f_{y,\ell}-\alpha_R} & 1 & \alpha_R & \dotsc & \alpha_R^{K+x}
    \end{bmatrix}
    \begin{bmatrix}
    W_{\theta,1}^{[\ell]}\\\vdots\\W_{\theta,y}^{[\ell]}\\v_0^{[\ell]}\\\vdots\\v_{K+x}^{[\ell]}
    \end{bmatrix}, \quad \ell\in\{1,\dotsc,K\},
\end{align}
where $v_i^{[\ell]}$ is the coefficient of $\alpha_n^i$ in $P_{\alpha_n}(K+x)$ of \eqref{last_ans}. Note that the $K\times y$ parameters in each subpacket of $W_{\theta}$ can be obtained by \eqref{mat} if $R=y+K+x+1$ is satsisfied. This determines the subpacketization (number of coded symbols in a subpacket of each submodel) as $y=R-K-x-1$, which results in the reading cost given by,
\begin{align} \label{R-cost}
    C_R=\frac{R\times K}{y\times K}=\frac{R}{R-K-x-1}.
\end{align}

\emph{Writing phase:} In the writing phase, the user sends $K$ symbols to each of the $R$ databases. The $\ell$th symbol (out of the $K$ symbols uploaded) is the combined update that consists of the updates of 
$W_{\theta}$ corresponding to the $\ell$th parameter of each of the $y$ coded parameters in each subpacket. The $K$ combined updates sent to database $n$, $n\in\{1,\dotsc,R\}$, are given by,
\begin{align}\label{update}
    U_{n,\ell}=\sum_{j=1}^y\prod_{i=1,i\neq j}^y(f_{i,\ell}-\alpha_n)\Tilde{\Delta}_{\theta,j}^{[\ell]}+\prod_{i=1}^y(f_{i,\ell}-\alpha_n)\hat{z}_{\ell}, \quad \ell\in\{1,\dotsc,K\},
\end{align}
where $\Tilde{\Delta}_{\theta,j}^{[\ell]}=\frac{\prod_{i=1,i\neq \ell}^K (f_{j,i}-f_{j,\ell})}{\prod_{i=1,i\neq j}^y (f_{i,\ell}-f_{j,\ell})}\Delta_{\theta,j}^{[\ell]}$ for $j\in\{1,\dotsc,y\}$, with $\Delta_{\theta,j}^{[\ell]}$ being the update of the $\ell$th parameter of the $j$th coded bit of the considered subpacket in submodel $\theta$ and $\hat{z}_{\ell}$ is a random noise symbol. Once database $n$ receives the update bits, it calculates the incremental update with the aid of the two matrices given by,
\begin{align}
    \Omega_{n,\ell}&=\diag\left(\frac{\prod_{r\in\mathcal{F}} (\alpha_r-\alpha_n)}{\prod_{r\in\mathcal{F}} (\alpha_r-f_{1,\ell})}1_M,\dotsc,\frac{\prod_{r\in\mathcal{F}} (\alpha_r-\alpha_n)}{\prod_{r\in\mathcal{F}} (\alpha_r-f_{y,\ell})}1_M\right),\\
    \Tilde{D}_{n,\ell}&=\diag\left(\frac{1}{\prod_{i=1}^K(f_{1,i}-\alpha_n)}1_M,\dotsc,\frac{1}{\prod_{i=1}^K(f_{y,i}-\alpha_n)}1_M\right),
\end{align}
where $\Omega_{n,\ell}$ is the null shaper in \cite{dropout} with $\mathcal{F}$ being any subset of randomly chosen databases satisfying $|\mathcal{F}|=x-y$. The null shaper is used to place some of the zeros ($x-y$ zeros) of the incremental update polynomial at specific $\alpha_n$s to reduce the writing cost by not having to send any updates to the databases corresponding to those $\alpha_n$s. Here, $\mathcal{F}$ is any subset of $x-y$ databases. $\Tilde{D}_{n,\ell}$ is a scaling matrix and $1_M$ is the vector of all ones of size $1\times M$. The incremental update is calculated as,
\begin{align}
    \bar{U}_{n,\ell}&=\Omega_{n,\ell}\times U_{n,\ell}\times \Tilde{D}_{n,\ell}\times Q_{n,\ell}, \quad \ell\in\{1,\dotsc,K\}\\
    &=\Omega_{n,\ell}\times U_{n,\ell}\times \begin{bmatrix}
    \frac{1}{\prod_{i=1}^K(f_{1,i}-\alpha_n)}\left(\frac{\prod_{i=1,i\neq \ell}^{K}(f_{1,i}-\alpha_n)}{\prod_{i=1,i\neq \ell}^{K}(f_{1,i}-f_{1,\ell})}e_M(\theta)+\prod_{i=1}^K (f_{1,i}-\alpha_n)\tilde{Z}_{1,\ell}\right)\\
    \vdots\\
    \frac{1}{\prod_{i=1}^K(f_{y,i}-\alpha_n)}\left(\frac{\prod_{i=1,i\neq \ell}^{K}(f_{y,i}-\alpha_n)}{\prod_{i=1,i\neq \ell}^{K}(f_{y,i}-f_{y,\ell})}e_M(\theta)+\prod_{i=1}^K (f_{y,i}-\alpha_n)\Tilde{Z}_{y,\ell}\right)
\end{bmatrix}\\ 
&=\Omega_{n,\ell}\times U_{n,\ell}\times \begin{bmatrix}
    \frac{1}{\prod_{i=1,i\neq \ell}^{K}(f_{1,i}-f_{1,\ell})}\frac{1}{(f_{1,\ell}-\alpha_n)}e_M(\theta)+\tilde{Z}_{1,\ell}\\
    \vdots\\
    \frac{1}{\prod_{i=1,i\neq \ell}^{K}(f_{y,i}-f_{y,\ell})}\frac{1}{(f_{y,\ell}-\alpha_n)}e_M(\theta)+\Tilde{Z}_{y,\ell}
\end{bmatrix}\\ 
&=\Omega_{n,\ell}\times \begin{bmatrix}
    \frac{1}{(f_{1,\ell}-\alpha_n)}\Delta_{\theta,1}^{[\ell]}e_M(\theta)+P_{\alpha_n}^{[1]}(y)\\
    \vdots\\
    \frac{1}{(f_{y,\ell}-\alpha_n)}\Delta_{\theta,y}^{[\ell]}e_M(\theta)+P_{\alpha_n}^{[y]}(y)
\end{bmatrix}\label{lemmm1}\\
    &=\begin{bmatrix}
        \frac{1}{f_{1,\ell}-\alpha_n}\Delta_{\theta,1}^{[\ell]}e_M(\theta)+P_{\alpha_n}^{[1]}(x)\\
        \vdots\\
        \frac{1}{f_{y,\ell}-\alpha_n}\Delta_{\theta,y}^{[\ell]}e_M(\theta)+P_{\alpha_n}^{[y]}(x)
    \end{bmatrix}, \quad \ell\in\{1,\dotsc,K\},\label{lemmm2}
\end{align}
where $P_{\alpha_n}^{[j]}(h)$ here is a vector of size $M\times1$ consisting of polynomials in $\alpha_n$ of degree $h$. Note that \eqref{lemmm1} and \eqref{lemmm2} are obtained by applying \cite[Lem.~1]{pruw_jpurnal} and \cite[Lem.~2]{pruw_jpurnal}, respectively. Since the incremental updates are of the same form as the storage in \eqref{storage}, the submodels are updated by,
\begin{align}
    S_n(t)=S_n(t-1)+\sum_{\ell=1}^K \bar{U}_{n,\ell}.    
\end{align}
The resulting writing cost is,
\begin{align} \label{W-cost}
    C_W=\frac{K\times(R-(x-y))}{K\times y}=\frac{2R-2x-K-1}{R-x-K-1},
\end{align}
which together with the reading cost in \eqref{R-cost} gives the total cost,
\begin{align}
    C_T=C_R+C_W=\frac{3R-2x-K-1}{R-x-K-1}.
\end{align}

The total cost is an increasing function of $x$ since $\frac{dC_T}{dx}=\frac{R+K+1}{(R-x-K-1)^2}>0$. Note that $x\geq y$ must be satisfied by $x$ in order to write to $y$ parameters using a single symbol. This is because the decomposition of the combined update in \eqref{lemmm1} results in a noise polynomial of degree $y$, which requires the existing storage to have at least a degree $y$ noise polynomial, as the incremental update is added to the existing storage in the updating process. Therefore, the optimum value of $x$ that minimizes the total cost is,
\begin{align}\label{sub_noise}
    x=\begin{cases}
        y=\frac{R-K-1}{2}, \quad & \text{if $R-K$ is odd},\\
        y+1=\frac{R-K}{2}, \quad & \text{if $R-K$ is even}.
\end{cases}
\end{align}
The resulting minimum total cost of the PRUW process with $(K,R)$ MDS coded storage is,
\begin{align}\label{totalcost2}
    C_T=\begin{cases}
    \frac{4R}{R-K-1}, \quad & \text{if $R-K$ is odd},\\
    \frac{4R-2}{R-K-2}, \quad & \text{if $R-K$ is even}.
    \end{cases}
\end{align}
Note that since the subpacketization $y\geq1$, $R$ and $K$ must satisfy,
\begin{align}\label{range}
    1\leq K\leq\begin{cases}
    R-3, \quad & \text{if $R-K$ is odd},\\
    R-4, \quad & \text{if $R-K$ is even}.
    \end{cases}
\end{align}

\subsection{Storage Mechanism}\label{general}

The proposed storage mechanism consists of submodel partitioning and submodel encoding, where the parameters are encoded with a specific $(K,R)$ MDS code and stored at different subsets of databases in parts. In this section, we find the optimum coding parameters and fractions of submodels stored in each database to minimize the total cost.

\subsubsection{Submodel Partitioning}

This determines the fractions of all $(K,R)$ MDS coded submodels to be stored in the $N$ databases with arbitrary storage constraints. Based on the $(K,R)$ MDS coded structure, each coded parameter must be stored in exactly $R$ databases. Let $\eta_i$ be the fraction of all submodels that are stored in the same subset of $R$ databases (the subset of databases indexed by $i$). Let $\mathcal{B}$ be the basis containing all $N\times1$ vectors with elements in $\{0,1\}$ with exactly $R$ ones, denoted by $\mathcal{B}=\{b_1,b_2,\dotsc,b_{|\mathcal{B}|}\}$.
Note that each $b_i$ corresponds to a specific subset of $R$ databases, indexed by $i$. Then, it is required to find the $\eta_i$s that satisfy,
\begin{align}
    \frac{1}{K}\sum_{i=1}^{|\mathcal{B}|}\eta_i b_i&=\mu, \quad \text{where} \quad  \mu=[\mu(1),\dotsc,\mu(N)]^T\label{p1}\\
    \sum_{i=1}^{|\mathcal{B}|}\eta_i&=1,\quad 
    0\leq\eta_1,\dotsc,\eta_{|\mathcal{B}|}\leq1, \label{p3}
\end{align}
to replicate each coded parameter at exactly $R$ databases, while ensuring that all databases are completely filled. The solution to this problem with uncoded parameters ($K=1$) is provided in \cite{heteroPIR} and \cite{utah} along with a necessary and sufficient condition for a solution to exist, given by,
\begin{align}\label{condition}
    \mu(n)\leq\frac{\sum_{n=1}^N\mu(n)}{R}, \quad n\in\{1,\dotsc,N\}.
\end{align}
The intuition behind this condition is as follows. Consider database $n$ which can store up to $\mu(n)ML$ bits. If each bit is expected to be replicated at $R$ databases, the total available storage in all databases $\sum_{i=1}^N\mu(n)ML$ must be greater than or equal to $\mu(n)MLR$, to successfully replicate the $\mu(n)ML$ bits of database $n$ in $R$ databases. This must be satisfied by all $N$ databases, which results in \eqref{condition}. Note that this condition is valid for any $(K,R)$ MDS coded setting, as the available storage in each database does not change with the submodel encoding structure.

In this work, we find the optimum coding parameters $K$ and $R$ that result in the minimum total cost of the PRUW process while satisfying \eqref{condition}, and solve \eqref{p1}-\eqref{p3} to find the partitions $\eta_i$ of coded submodels to be stored in each database. 

\subsubsection{Submodel Encoding}\label{encode}

For a given set of storage constraints $\{\mu(n)\}_{n=1}^N$, the total available storage is $p=\sum_{n=1}^N \mu(n)$ where the intuition behind the value of $p$ is the maximum number of times each parameter in the model can be replicated in the system of databases if the parameters are uncoded. Let the parameters of the model be stored using a $(K,R)$ MDS code. Therefore, the total number of coded parameters in the model is $\frac{ML}{K}$, which allows each coded parameter to be replicated at a maximum of,
\begin{align}\label{repl}
    R\leq\frac{\sum_{n=1}^N\mu(n)ML}{\frac{ML}{K}}=Kp
\end{align}
databases, given that the value of $K$ satisfies $K\leq\frac{N}{p}$. At the same time, to completely utilize the available space in each database, the number of parameters stored in each database must satisfy,
\begin{align}
    \mu(n)ML\leq\frac{ML}{K}, \quad n\in\{1,\dotsc,N\} \ \implies \ K\leq\frac{1}{\Bar{\mu}}, 
\end{align}
where $\Bar{\mu}=\max_{n}\mu(n)$. Therefore, the coding parameter $K$ must satisfy,
\begin{align}\label{condK}
    K\leq\min\left\{\frac{1}{\Bar{\mu}},\frac{N}{p}\right\}=\frac{1}{\Bar{\mu}},
\end{align}
as $p=\sum_{n=1}^N \mu(n)\leq\Bar{\mu}N$ implies $\frac{1}{\Bar{\mu}}\leq \frac{N}{p}$. Since the total cost of the PRUW scheme decreases with the number of replications $R$ (see \eqref{totalcost2}), for given $K,p\in\mathbb{Z}^+$ satisfying \eqref{condK}, the optimum $R$ for the $(K,R)$ MDS code is given by $Kp$ (from \eqref{repl}), and the resulting total cost is,
\begin{align}
    C_T(K,R)=\begin{cases}
    \frac{4Kp}{Kp-K-1}, \quad & \text{odd $K$, even $p$},\\
    \frac{4Kp-2}{Kp-K-2}, \quad & \text{otherwise},
    \end{cases}
\end{align}
which decreases with $K$. Therefore, for a given set of storage constraints $\{\mu(n)\}_{n=1}^N$, the minimum total cost is achieved by the $(K,R)$ MDS code in \eqref{storage} with $K$ and $R$ given by,
\begin{align}\label{opt}
    K=\frac{1}{\Bar{\mu}}\stackrel{\text{def}}{=} k, \qquad \quad  R=kp\stackrel{\text{def}}{=} r
\end{align}
which automatically satisfies \eqref{condition} since
$\frac{\sum_{n=1}^N\mu(n)}{r}\!=\!\Bar{\mu}\!\geq\!\mu(n)$ for all $n$. However, since $k$ and $r$ are not necessarily integers, we need additional calculations to obtain the optimum $(k,r)$ MDS codes with $k,r\in\mathbb{Z^+}$ that collectively result in the lowest total cost. 

Consider the general case where $k, r\notin\mathbb{Z}^+$. The general idea here is to divide each submodel into four sections and encode them using the four MDS codes given in Table~\ref{codes}. The sizes (fractions) of the four sections are defined by the parameters $\alpha$, $\beta$ and $\delta$, where $0\leq \alpha,\beta,\delta\leq1$ (see Table~\ref{codes}). The total spaces allocated for the four MDS codes in the entire system of databases are given by,
\begin{align}
    \sum_{n=1}^N\hat{\mu}_1(n)&=\frac{\alpha\beta}{\lfloor k\rfloor}\lfloor r\rfloor\label{eq1}\\
    \sum_{n=1}^N\hat{\mu}_2(n)&=\frac{\alpha(1-\beta)}{\lfloor k\rfloor}\lceil r\rceil\label{eq2}\\
    \sum_{n=1}^N\bar{\mu}_1(n)&=\frac{(1-\alpha)\delta}{\lceil k\rceil}\lfloor r\rfloor\label{eq3}\\ 
    \sum_{n=1}^N\bar{\mu}_2(n)&=\frac{(1-\alpha)(1-\delta)}{\lceil k\rceil}\lceil r\rceil\label{eq4}
\end{align}

\begin{table}[ht]
\begin{center}
\begin{tabular}{ |c|c|c|c| }
\hline
  case & MDS code & fraction of submodel & space allocated in database $n$\\ 
  \hline
  \text{1} & $(\lfloor k\rfloor, \lfloor r\rfloor ) $ & $\alpha\beta$ & $\hat{\mu}_1(n)$\\
  \hline
  \text{2} & $(\lfloor k\rfloor, \lceil r\rceil ) $ & $\alpha(1-\beta)$& $\hat{\mu}_2(n)$\\
  \hline
  \text{3} & $(\lceil k\rceil, \lfloor r\rfloor )$ & $(1-\alpha)\delta$& $\bar{\mu}_1(n)$\\
  \hline
  \text{4} & $(\lceil k\rceil, \lceil r\rceil )$  & $(1-\alpha)(1-\delta)$& $\bar{\mu}_2(n)$\\  
  \hline
\end{tabular}
\end{center}
\caption{Fractions of submodels and corresponding MDS codes.}
\label{codes}
\end{table}

The space allocated for each individual MDS code in each database must satisfy \eqref{condition} separately, to ensure that all databases are completely filled while also replicating each coded parameter at the respective number of databases, i.e.,
\begin{align}
    \hat{\mu}_1(n)&\leq\frac{\alpha\beta}{\lfloor k\rfloor}\label{neq1}\\
    \hat{\mu}_2(n)&\leq\frac{\alpha(1-\beta)}{\lfloor k\rfloor}\label{neq2}\\
    \bar{\mu}_1(n)&\leq\frac{(1-\alpha)\delta}{\lceil k\rceil}\label{neq3}\\
    \bar{\mu}_2(n)&\leq\frac{(1-\alpha)(1-\delta)}{\lceil k\rceil}.\label{neq4}
\end{align}
All four storage allocations in each database must satisfy,  
\begin{align}\label{sum}
\hat{\mu}_1(n)+\hat{\mu}_2(n)+\bar{\mu}_1(n)+\bar{\mu}_2(n)=\mu(n), \quad \forall n.
\end{align}
It remains to find the values of $\hat{\mu}_1(n)$, $\hat{\mu}_2(n)$, $\bar{\mu}_1(n)$, $\bar{\mu}_2(n)$, $\alpha$, $\beta$ and $\delta$ that minimize the total cost given by,
\begin{align}\label{min}
    C&=\alpha\beta C_T(\lfloor k\rfloor,\lfloor r\rfloor)+\alpha(1-\beta) C_T(\lfloor k\rfloor,\lceil r\rceil)+(1-\alpha)\delta C_T(\lceil k\rceil,\lfloor r\rfloor)\nonumber\\
    &\quad+(1-\alpha)(1-\delta) C_T(\lceil k\rceil,\lceil r\rceil)
\end{align}
while satisfying \eqref{eq1}-\eqref{sum}, where $C_T$ is defined in \eqref{totalcost}. We consider two cases, 1) $\alpha=1$, 2) $\alpha<1$. When $\alpha=1$, only cases 1, 2 in Table~\ref{codes} are used in storage, and \eqref{condition} becomes $R\leq\lfloor k\rfloor p$, which results in $\lfloor k\rfloor p=s$ maximum replications. This replaces all $r$s in Table~\ref{codes} and \eqref{eq1}-\eqref{neq4} by $s$ when $\alpha=1$. The optimum values of the parameters in Table~\ref{codes} that minimize the total cost in \eqref{min} when $\alpha=1$ and $\alpha<1$ are stated in Lemma~\ref{lem0} and Lemmas~\ref{lem1}-\ref{lem2}, respectively. For a given set of storage constraints, we choose the set of optimum values corresponding to either $\alpha=1$ or $\alpha<1$, based on the case that results in the minimum total cost. In other words, $C_1$ and $C_2$ stated in \eqref{c1} and \eqref{c2} in Section~\ref{main} indicate the total costs corresponding to the two cases $\alpha=1$ and $\alpha<1$, respectively, from which the minimum is chosen.

\begin{lemma}\label{lem0}
When $\alpha=1$, $\beta$ is fixed at $\beta=\lceil s\rceil-s$ to satisfy \eqref{eq1}-\eqref{eq2}, and $\hat{\mu}_1(n),\hat{\mu}_2(n)$ satisfying \eqref{neq1}-\eqref{neq2} are given by,
\begin{align}
    \hat{\mu}_1(n)&=\Tilde{m}(n)+(\mu(n)-\Tilde{m}(n)-\Tilde{h}(n))\Tilde{\gamma}\label{mu11}\\
    \hat{\mu}_2(n)&=\Tilde{h}(n)+(\mu(n)-\Tilde{m}(n)-\Tilde{h}(n))(1-\Tilde{\gamma})\label{mu21}
\end{align}
for all $n\in\{1,\dotsc,N\}$ where,
\begin{align}
    &\Tilde{m}(n)\!=\!\left[\mu(n)\!-\!\frac{s\!-\!\lfloor s\rfloor}{\lfloor k\rfloor}\right]^+\!\!\!\!,\quad \Tilde{h}(n)\!=\!\left[\mu(n)\!-\!\frac{\lceil s\rceil\!-\!s}{\lfloor k\rfloor}\right]^+\label{tildemh}\\
    &\Tilde{\gamma}=\frac{\frac{\lfloor s\rfloor}{\lfloor k\rfloor}(\lceil s\rceil-s)-\sum_{n=1}^N \Tilde{m}(n)}{p-\sum_{n=1}^N \Tilde{m}(n)-\sum_{n=1}^N \Tilde{h}(n)},\label{tildegamma}
\end{align}
with $[x]^+\!=\!\max\{x,0\}$. The resulting total cost is given in \eqref{c1}.
\end{lemma}

\begin{lemma}\label{lem1}
The following values of $\hat{\mu}_1(n)$, $\hat{\mu}_2(n)$, $\bar{\mu}_1(n)$ and $\bar{\mu}_2(n)$ for $n\in\{1,\dotsc,N\}$ satisfy \eqref{eq1}-\eqref{sum} with any $\alpha<1$, $\beta$ and $\delta$ that satisfy $\alpha\geq\frac{\lfloor k\rfloor}{k}(\lceil k\rceil-k)$, $\beta\geq\left[1-\frac{\lfloor k\rfloor}{k\alpha}(r-\lfloor r\rfloor)\right]^+$ and $\delta\geq\left[1-\frac{\lceil k\rceil}{k(1-\alpha)}(r-\lfloor r\rfloor)\right]^+$,
\begin{align}
    \hat{\mu}_1(n)&=\begin{cases}
    \hat{\mu}(n)\beta, & \text{if $\beta\in\{0,1\}$}\\
    \hat{m}(n)+(\hat{\mu}(n)-\hat{m}(n)-\hat{h}(n))\hat{\gamma}, & \text{if $\beta\in(0,1)$}
    \end{cases}\label{e1}\\
    \hat{\mu}_2(n)&=\begin{cases}
    \hat{\mu}(n)(1-\beta), & \text{if $\beta\in\{0,1\}$}\\
    \hat{h}(n)+(\hat{\mu}(n)-\hat{m}(n)-\hat{h}(n))(1-\hat{\gamma}), & \text{if $\beta\in(0,1)$}
    \end{cases}\label{muhat2}\\
    \bar{\mu}_1(n)&=\begin{cases}
    \bar{\mu}(n)\delta, & \text{if $\delta\in\{0,1\}$}\\
    \bar{m}(n)+(\bar{\mu}(n)-\bar{m}(n)-\bar{h}(n))\bar{\gamma}, & \text{if $\delta\in(0,1)$}
    \end{cases}\label{mubar1}\\
    \bar{\mu}_2(n)&=\begin{cases}
    \bar{\mu}(n)(1-\delta), & \text{if $\delta\in\{0,1\}$}\\
    \bar{h}(n)+(\bar{\mu}(n)-\bar{m}(n)-\bar{h}(n))(1-\bar{\gamma}), & \text{if $\delta\in(0,1)$}
    \end{cases}\label{e2}   
\end{align}
for all $n\in\{1,\dotsc,N\}$ where,
\begin{align}
    \hat{\mu}(n)&=m(n)+(\mu(n)-m(n)-h(n))\gamma\label{ee1}\\
    \bar{\mu}(n)&=h(n)+(\mu(n)-m(n)-h(n))(1-\gamma)\label{eee1}\\
    m(n)&\!=\!\left[\mu(n)\!-\!\frac{1-\alpha}{\lceil k\rceil}\right]^+,\quad h(n)\!=\!\left[\mu(n)\!-\!\frac{\alpha}{\lfloor k\rfloor}\right]^+\label{mh}\\    
    \gamma&=\frac{\frac{\alpha}{\lfloor k\rfloor}(\lceil r\rceil-\beta)-\sum_{n=1}^N m(n)}{p-\sum_{n=1}^N m(n)-\sum_{n=1}^N h(n)}\label{gama}\\
    \hat{m}(n)&\!=\!\left[\hat{\mu}(n)\!-\!\frac{\alpha(1-\beta)}{\lfloor k\rfloor}\right]^+\!\!,\quad\!\! \hat{h}(n)\!=\!\left[\hat{\mu}(n)\!-\!\frac{\alpha\beta}{\lfloor k\rfloor}\right]^+\\
    \hat{\gamma}&=\frac{\frac{\alpha\beta}{\lfloor k\rfloor}\lfloor r\rfloor-\sum_{n=1}^N \hat{m}(n)}{\frac{\alpha}{\lfloor k\rfloor}(\lceil r\rceil-\beta)-\sum_{n=1}^N \hat{m}(n)-\sum_{n=1}^N \hat{h}(n)}\label{gamahat}\\
    \bar{m}(n)&\!=\!\left[\bar{\mu}(n)\!-\!\frac{(1\!-\!\alpha)(1\!-\!\delta)}{\lceil k\rceil}\right]^+\!\!\!\!,\quad\!\! \bar{h}(n)\!=\!\left[\bar{\mu}(n)\!-\!\frac{(1\!-\!\alpha)\delta}{\lceil k\rceil}\right]^+\\
    \bar{\gamma}&=\frac{\frac{(1-\alpha)\delta}{\lceil k\rceil}\lfloor r\rfloor-\sum_{n=1}^N \bar{m}(n)}{\frac{1-\alpha}{\lceil k\rceil}(\lceil r\rceil-\delta)-\sum_{n=1}^N \bar{m}(n)-\sum_{n=1}^N \bar{h}(n)}.\label{ee2}
\end{align}
\end{lemma}

\begin{lemma}\label{lem2}
For the case where $\alpha<1$, the values of $\alpha$, $\beta$ and $\delta$ that minimize the total cost in \eqref{min} while satisfying the constraints in Lemma~\ref{lem1} are specified in Theorem~\ref{thm1}, and the corresponding total cost is given in \eqref{c2}.
\end{lemma}

The three lemmas stated above provide the proof of Theorem~\ref{thm1}. The proofs of Lemmas~\ref{lem0},~\ref{lem1} and~\ref{lem2} are given in the Appendix. Once all parameters in Table~\ref{codes} are determined from Lemmas~\ref{lem0},~\ref{lem1},~\ref{lem2}, equations \eqref{p1}-\eqref{p3} are solved for $\eta_i$s for the four sets of storage allocations, $\{\hat{\mu}_1\}_{n=1}^N$, $\{\hat{\mu}_2\}_{n=1}^N$, $\{\bar{\mu}_1\}_{n=1}^N$, $\{\bar{\mu}_2\}_{n=1}^N$ separately. These four sets of solutions determine where each individual coded symbol is replicated in the system of databases. Then, the storage structure and the scheme provided in Section~\ref{pruw-general} is used to perform private FSL.

\subsection{Example}

As an example, consider a PRUW setting with $N=12$ databases, where $\mu(1)=\dotsc\mu(5)=0.37$ and $\mu(6)=\dotsc\mu(12)=0.35$. Note that the values of $k$, $r$ and $s$ here are the same as what is considered in Remark~\ref{rem1}, and therefore result in the same $\alpha$, $\beta$, $\delta$ and total cost stated in Remark~\ref{rem1}, i.e., $k=2.7$, $r=11.61$, $s=8.6$, $\alpha=\frac{2}{9}$, $\beta=1$ and $\delta=\frac{9}{70}$. Using \eqref{e1}-\eqref{ee2} we find the storage allocations as,
\begin{align}
    \hat{\mu}_1(n)&=\hat{\mu}(n)=\begin{cases}
        0.1107, & \text{for $n=1,\dotsc5$}\\
        0.0951, & \text{for $n=6,\dotsc12$}
    \end{cases}\label{dif1}\\
    \Bar{\mu}_1(n)&=\begin{cases}
        0.033, & \text{for $n=1,\dotsc5$}\\
        0.029, & \text{for $n=6,\dotsc12$}
    \end{cases}\label{dif2}\\
    \hat{\mu}_2(n)&=0,\quad \Bar{\mu}_2(n)=0.226,\quad \forall n.\label{same}
\end{align}
Note that the $(\lfloor k\rfloor, \lceil r\rceil)$ MDS code is not used as $\hat{\mu}_2(n)=0$, $\forall n$, and the allocated space for the  $(\lceil k\rceil, \lceil r\rceil)$ MDS code in all databases is the same since $\bar{\mu}_2(n)=0.226$, $\forall n$. Therefore, the optimum storage mechanism for PRUW with homogeneous storage constraints presented in Section~\ref{improve} is used to place the $(\lceil k\rceil, \lceil r\rceil)$ coded parameters. The storage allocations for $(\lfloor k\rfloor, \lfloor r\rfloor)$ and $(\lceil k\rceil, \lfloor r\rfloor)$ codes have different storage allocations for different databases (two different values in \eqref{dif1}, \eqref{dif2}). Therefore, for these two cases, we solve \eqref{p1}-\eqref{p3} separately, to find the fractions of coded parameters to be stored in each database. For example, for the $(\lfloor k\rfloor, \lfloor r\rfloor)$ MDS code, we find the solution (values of $\eta_i$) to \eqref{p1}-\eqref{p3} with $N=12$, $K=\lfloor k\rfloor=2$, $R=\lfloor r\rfloor=11$ and $\mu=[\hat{\mu}_1(1):\hat{\mu}_1(12)]^T$ as follows. For this case, \eqref{p1}-\eqref{p3} is given by,
\begin{align}
    \sum_{i=1}^{12}\frac{\alpha\beta ML}{\lfloor k\rfloor}\eta_ib_i&=\mu ML\\
    \sum_{i=1}\eta_i&=1,
\end{align}
where $b_i$ is the all ones vector of size $12\times1$ with a zero at the $i$th position. As a solution, we get $\tilde{\eta}_1=\dotsc=\tilde{\eta}_5=0$ and $\tilde{\eta}_6=\dotsc=\tilde{\eta}_{12}=0.0315$, where $\tilde{\eta}_i=\alpha\beta\eta_i$. To explain how the submodel parameters are stored, consider $\tilde{\eta}_6=0.0315$. From each submodel, a fraction of $\tilde{\eta}_6=0.0315$ is chosen to be replicated at all databases except for database 6. The chosen set of parameters from all submodels are $(\lfloor k\rfloor,\lfloor r\rfloor)$ MDS coded based on the structure shown in Section~\ref{pruw-general}, and the PRUW process is carried out accordingly.

\subsection{Improved Scheme for Special Cases}

The key idea behind the scheme proposed in Section~\ref{general} is to find the optimum linear combination of $(K,R)$ MDS codes to store the submodel parameters, as shown in the example in Fig.~\ref{example}. However, note in the same figure that the achievable total communication costs have a fluctuating structure based on whether the values of $K$ and $R$ result in odd or even $R-K$ in \eqref{totalcost2}. For any $(K,R)$ MDS code with even $R-K$, the total cost results in a \emph{local peak}. Therefore, the total communication cost can be decreased further if for a given set of storage constraints $\{\mu(n)\}_{n=1}^N$, the model parameters can be stored as a linear combination of $(K,R)$ MDS codes with only odd $R-K$ instead of the four codes considered in Table~\ref{codes}. This eliminates the involvement of the \emph{local peaks}. We begin the analysis of such cases with the following lemma.

\begin{lemma}\label{lemm1}
Let $(\boldsymbol{\mu_1},C_T(\boldsymbol{\mu_1}))$ and $(\boldsymbol{\mu_2},C_T(\boldsymbol{\mu_2}))$ be two pairs of storage constraints and the corresponding achievable total costs. The storage constraints are given by $\boldsymbol{\mu_1}=\{\mu_1(n)\}_{n=1}^N$ and $\boldsymbol{\mu_2}=\{\mu_2(n)\}_{n=1}^N$. Then, the pair $(\boldsymbol{\mu},C_T(\boldsymbol{\mu}))$ is also achievable for any $\gamma\in[0,1]$ where $\boldsymbol{\mu}=\{\mu(n)\}_{n=1}^N$ with,
\begin{align}
    \mu(n)&=\gamma\mu_1(n)+(1-\gamma)\mu_2(n),\quad n\in\{1,\dotsc,N\}\\
    C_T(\boldsymbol{\mu})&=\gamma C_T(\boldsymbol{\mu_1})+(1-\gamma)C_T(\boldsymbol{\mu_2}) 
\end{align}
\end{lemma}

\begin{Proof}
Since $(\boldsymbol{\mu_1},C_T(\boldsymbol{\mu_1}))$ and $(\boldsymbol{\mu_2},C_T(\boldsymbol{\mu_2}))$ are achievable, let $S_1$ and $S_2$ be the schemes that produce the achievable pairs $(\boldsymbol{\mu_1},C_T(\boldsymbol{\mu_1}))$ and $(\boldsymbol{\mu_2},C_T(\boldsymbol{\mu_2}))$, respectively. A new scheme can be generated by applying $S_1$ on a $\gamma$ fraction of bits of all submodels and $S_2$ on the rest of the bits. The storage capacity of database $n$ in this combined scheme is given by $\gamma ML\mu_1(n)+(1-\gamma)ML\mu_2(n)=\mu(n) ML$ bits. The corresponding total cost is, 
\begin{align}
    C_T=\frac{\gamma LC_T(\boldsymbol{\mu_1})+(1-\gamma)LC_T(\boldsymbol{\mu_2})}{L}=\gamma C_T(\boldsymbol{\mu_1})+(1-\gamma)C_T(\boldsymbol{\mu_2}).
\end{align}
completing the proof.
\end{Proof}

\begin{corollary}
    Consider a PRUW setting with an arbitrary set of storage constraints $\boldsymbol{\mu}=\{\mu(n)\}_{n=1}^N$. Based on Lemma~\ref{lemm1}, a $\gamma$ fraction of all submodel parameters can be stored using any $(K_1,R_1)$ MDS code, and the rest of the $1-\gamma$ fraction can be stored using any $(K_2,R_2)$ MDS code 
    to achieve a total cost of $\gamma C_T(K_1,R_1)+(1-\gamma) C_T(K_2,R_2)$ if there exist two sets of storage constraints $\boldsymbol{\mu_1}=\{\mu_1(n)\}_{n=1}^N$ and $\boldsymbol{\mu_2}=\{\mu_2(n)\}_{n=1}^N$ that only use $(K_1,R_1)$ and $(K_2,R_2)$ MDS codes, respectively, to store the submodel parameters. For this, $\boldsymbol{\mu_1}$ and $\boldsymbol{\mu_2}$ must satisfy the following conditions.  
\begin{align}\label{eqfirst}
    \gamma\mu_1(n)+(1-\gamma)\mu_2(n)&=\mu(n),\quad n\in\{1,\dotsc,N\}\\
    \sum_{n=1}^N \mu_1(n)&=\frac{R_1}{K_1}\label{eqsec}\\
    \max_{n} \mu_1(n)&\leq\frac{1}{K_1}\label{eqth}\\
    \sum_{n=1}^N \mu_2(n)&=\frac{R_2}{K_2}\label{eqfo}\\
    \max_{n} \mu_2(n)&\leq\frac{1}{K_2}\label{eqlast}
\end{align}
where \eqref{eqfirst} is straightforward from Lemma~\ref{lemm1}, \eqref{eqsec}, \eqref{eqfo} are based on the fact that a $(K,R)$ MDS code combines $K$ bits together and stores them at $R$ databases, resulting in a total of $\frac{ML}{K}\times R$ bits stored across all databases, and \eqref{eqth}, \eqref{eqlast} are required by \eqref{condition}. 
\end{corollary}

The method proposed in Section~\ref{general} cannot be used to find $\boldsymbol{\mu_1}=\{\mu_1(n)\}_{n=1}^N$ and $\boldsymbol{\mu_2}=\{\mu_2(n)\}_{n=1}^N$ in \eqref{eqfirst}-\eqref{eqlast} for any given set of arbitrary storage constraints because of \eqref{eqfirst} and limitations on $R_1,R_2,K_1,K_2$. In this section, we discuss a specific type of storage constraints $\boldsymbol{\mu}=\{\mu(n)\}_{n=1}^N$ for which a direct solution to \eqref{eqfirst}-\eqref{eqlast} is available. For this type of storage constraints, the \emph{local peaks} in the achievable costs curves can be eliminated, which results in reduced total communication costs compared to what can be achieved from the scheme in Section~\ref{general} for the same storage constraints.

Consider the case of homogeneous storage constraints, i.e., $\mu(n)=\mu$ for $n\in\{1,\dotsc,N\}$. This is the specific type of storage constraints we consider in this section with a direct solution to \eqref{eqfirst}-\eqref{eqlast}. Note that for this case, \eqref{eqfirst}, \eqref{eqsec} and \eqref{eqfo} imply,
\begin{align}\label{homo}
    \mu=\gamma\frac{R_1}{NK_1}+(1-\gamma)\frac{R_2}{NK_2}.
\end{align}
Therefore, for any $\mu$ such that $\frac{R_1}{NK_1}\leq\mu\leq\frac{R_2}{NK_2},$\footnote{Without loss of generality, we assume that $\frac{R_1}{K_1}\leq\frac{R_2}{K_2}$.} one solution to \eqref{eqfirst}-\eqref{eqlast} is given by,
\begin{align}
    \mu_1(n)&=\frac{R_1}{NK_1},\quad n\in\{1,\dotsc,N\}\\
    \mu_2(n)&=\frac{R_2}{NK_2}, \quad n\in\{1,\dotsc,N\},
\end{align}
as $N\geq R_1,R_2$ must be satisfied by any $(K_i,R_i)$ MDS coded storage in a system of $N$ databases. The value of $\gamma$ for the given $\mu$ is determined by \eqref{homo}.  

The above solution basically implies that when a given homogeneous storage constraint $\mu$ is in the range $\frac{R_1}{NK_1}\leq\mu\leq\frac{R_2}{NK_2}$ for any $R_1,R_2,K_1,K_2$, a total cost of $\gamma C_T(K_1,R_1)+(1-\gamma)C_T(K_2,R_2)$ is achievable, where $\gamma=\frac{N\mu-\frac{R_2}{K_2}}{\frac{R_1}{K_1}-\frac{R_2}{K_2}}$. This further implies that all points on the line connecting $(\frac{R_1}{NK_1},C_T(\frac{R_1}{NK_1}))$ and $(\frac{R_2}{NK_2},C_T(\frac{R_2}{NK_2}))$ for any $R_1,R_2,K_1,K_2$ with $\frac{R_1}{K_1}\leq\frac{R_2}{K_2}$ are achievable. This allows for any point on the line connecting the adjacent local minima in the total costs curves to be achievable, which eliminates the \emph{local peaks}. 

Next, we present the general storage scheme for homogeneous storage constraints based on the above arguments. 

\section{Proposed Scheme: Homogeneous Storage Constraints}\label{improve}

In this section, we present the general scheme for arbitrary homogeneous storage constraints denoted by $\mu(n)=\mu$, $n\in\{1,\dotsc,N\}$. The basic idea of this scheme is to find all achievable pairs of the form $(\mu,C_T(\mu))$,\footnote{$\boldsymbol{\mu}$ in Lemma~\ref{lemm1} is replaced by $\mu$, as all storage constraints are equal.}, find its lower convex hull, and apply Lemma~\ref{lemm1} for a given $\mu$ with the closest points on the convex hull  to obtain the minimum achievable cost with the PRUW scheme presented in Section~\ref{pruw-general}. The detailed storage scheme is given next.

For a given $N$ we first find the achievable pairs of $(\mu,C_T(\mu))$ as follows. Let $\mu=\frac{R}{NK_R}$ for $R=4,\dotsc,N$ and $K_R=1,\dotsc,R-3$. For a given $\mu$ with a given $R$ and $K_R$, the following steps need to be followed in order to perform PRUW while meeting the storage constraint: 
\begin{enumerate}
    \item Divide the $L$ bits of each submodel into $N$ sections and label them as $\{1,\dotsc,N\}$.
    \item Allocate sections $n:(n-1+R)\!\!\mod N$ to database $n$ for $n\in\{1,\dotsc,N\}$.\footnote{The indices here follow a cyclic pattern, i.e., if $(n-1+R)\!\!\mod N<n$, then $n:(n-1+R)\!\!\mod N$ implies $\{n,\dotsc,N,1,\dotsc,(n-1+R)\!\!\mod N\}$.}
    \item Use the storage specified in \eqref{storage} with $K=K_R$ and $x,y$ given in \eqref{sub_noise} to encode each of the allocated sections of all submodels. Note that a given coded bit of a given section of each submodel stored across different databases contains the same noise polynomial that only differs in $\alpha_n$.
    \item Use the PRUW scheme described in Section~\ref{pruw-general} on each of the subsets of $n:(n-1+R)\!\!\mod N$ databases to read/write to section $(n-1+R)\mod N$ of the required submodel for $n\in\{1,\dotsc,N\}$.
\end{enumerate}

For each $\mu=\frac{R}{NK_R}$, $R=4,\dotsc,N$, $K_R=1,\dotsc,R-3$, the above process encodes the submodel parameters using a $(K_R,R)$ MDS code, and gives an achievable $(\mu,C_T(\mu))$ pair, where $C_T(\mu)$ is given by
\begin{align}\label{Tmu}
    C_T(\mu)=\begin{cases}
        \frac{4R}{R-K_R-1}, & \text{if $R-K_R$ is odd}\\
        \frac{4R-2}{R-K_R-2}, & \text{if $R-K_R$ is even}.
    \end{cases}
\end{align}
Note that the above two cases,  which correspond to the value of $(R-K_R)\!\!\mod2$, are a result of two different schemes. The case with odd values of $R-K_R$ has a subpacketization that is equal to the degree of noise polynomial in storage, which does not require the null shaper, while the case with even values of $R-K_R$ contains two more noise terms than the subpacketization, which requires the null shaper; see \eqref{sub_noise}. The scheme corresponding to even values of $R-K_R$ is inefficient compared to the even case due to the additional noise term present in storage. This observation combined with Lemma~\ref{lemm1} results in the following lemma, which formally states how the \emph{local peaks} in achievable costs can be eliminated.

\begin{lemma}\label{lemm2}
For a given $\mu=\frac{R}{NK_R}$, if $R$ and $K_R$ are such that $R-K_R$ is even, it is more efficient to perform a linear combination of two PRUW schemes with nearest two odd $R^{[i]}-K_R^{[i]}$, $i=1,2$, instead of performing direct PRUW with the given $R$ and $K_R$, while satisfying the same storage constraint $\mu$, i.e., with $\mu_1=\frac{R^{[1]}}{NK_R^{[1]}}$ and $\mu_2=\frac{R^{[2]}}{NK_R^{[2]}}$.  
\end{lemma}

\begin{Proof}
For a given $\mu=\frac{R}{NK_R}$ with even $R-K_R$, the nearest $\mu_1=\frac{R^{[1]}}{NK_R^{[1]}}$ is $\frac{R-1}{NK_R}$, since \eqref{range} with $K$ replaced by $K_R$ needs to be satisfied for the PRUW scheme to work. Similarly, $\mu_2=\frac{R+1}{NK_R}$. Let $C_T(\mu)$, $C_T(\mu_1)$ and $C_T(\mu_2)$ be the total costs incurred by the scheme with $\mu$, $\mu_1$ and $\mu_2$, respectively. From \eqref{Tmu}, we have
\begin{align}
    C_T(\mu)=\frac{4R-2}{R-K_R-2}>C_T(\mu_1)=\frac{4R-4}{R-K_R-2}>C_T(\mu_2)=\frac{4(R+1)}{R-K_R}.
\end{align}
Note that $\mu_1<\mu<\mu_2$. From Lemma~\ref{lemm1}, there exists some $\gamma\in[0,1]$ that allocates the storage for the two PRUW schemes corresponding to $\mu_1$ and $\mu_2$ that achieves the same storage constraint as $\mu$, and results in a total cost of $\gamma C_T(\mu_1)+(1-\gamma)C_T(\mu_2)$, that satisfies
\begin{align}
    C_T(\mu_2)<\gamma C_T(\mu_1)+(1-\gamma)C_T(\mu_2)<C_T(\mu_1)<C_T(\mu),
\end{align}
completing the proof. 
\end{Proof}
Once the basic $(\mu,C_T(\mu))$ pairs corresponding to $\mu=\frac{R}{NK_R}$ for $R=4,\dotsc,N$, $K_R=1,\dotsc,R-3$ with odd $R-K_R$ are obtained, the minimum achievable total cost of the improved scheme for any $\mu$ is characterized by the lower convex hull of the above basic $(\mu,C_T(\mu))$ pairs, denoted by $T_{ach}$. This is straightforward from Lemmas~\ref{lemm1} and~\ref{lemm2}. Therefore, for a given $N$ and $\mu$, if $\mu_1=\frac{R_1}{NK_1}$ and $\mu_2=\frac{R_2}{NK_2}$ are the nearest storage constraints to $\mu$ such that $\mu_1\leq\mu\leq\mu_2$, with $(\mu_1,C_T(\mu_1))$ and $(\mu_1,C_T(\mu_1))$ being elements of the set of basic pairs that determine $T_{ach}$, a total cost of $\gamma C_T(K_1,R_1)+(1-\gamma)C_T(K_2,R_2)$ with $\gamma=\frac{\mu_2-\mu}{\mu_2-\mu_1}$ is achievable by storing $\gamma$ fraction of all submodels using a $(K_1,R_1)$ MDS code, and the rest of the $1-\gamma$ fractions of all submodels using a $(K_2,R_2)$ MDS code. Once the storage is determined, the PRUW scheme presented in Section~\ref{pruw-general} is used to perform the private FSL.

\subsection{Comparison with Other Schemes}

\begin{figure}[t]
    \centering
    \includegraphics[scale=0.9]{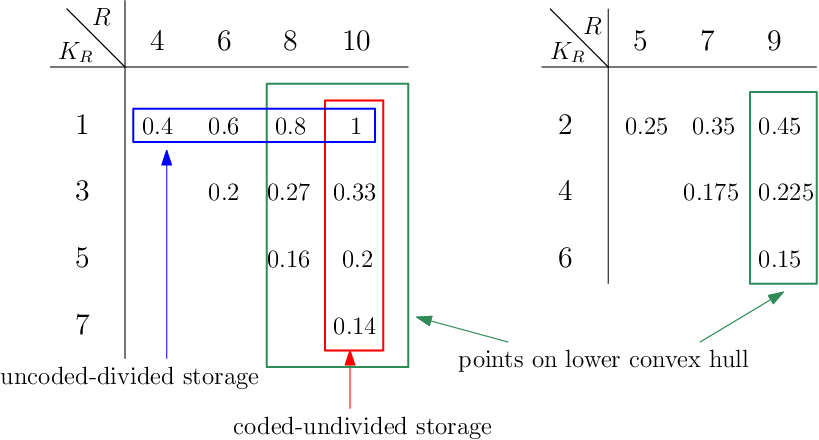}
    \caption{All possible pairs of $(R,K_R)$ and corresponding values of $\mu$ for $N=10$.}
    \label{fig2}
\end{figure}

The two straightforward methods to handle homogeneous storage constraints is to consider divided storage and coded storage. Divided storage, i.e., $K_R=1$, $R<N$ is where the submodel parameters are uncoded, but divided and stored at subsets of databases to meet the storage constraints. Coded storage, i.e., $K_R>1$, $R=N$ is where the submodel parameters are encoded to combine multiple symbols into a single symbol and stored at all $N$ databases. Note that both divided and coded storage mechanisms are subsets of the proposed storage mechanism which considers all cases that correspond to $K_R\geq1$ and $R\leq N$. In other words, the proposed scheme can be viewed as a hybrid mechanism of both divided and coded schemes. The hybrid scheme achieves lower total costs compared to divided and coded schemes, as it considers a larger set of basic $(\mu,C_T(\mu))$ pairs to find the lower convex hull, which includes all points considered in divided and coded schemes individually. As an illustration, consider the example with $N=10$ databases. The proposed hybrid storage mechanism first determines the basic achievable $\left(\mu=\frac{R}{NK_R},C_T(\mu)\right)$ pairs for $R=4,\dotsc,N$, $K_R=1,\dotsc,R-3$ with odd $R-K_R$. The pairs of $(R,K_R)$ and the corresponding values of $\mu$ are shown in Fig.~\ref{fig2}. 

\begin{figure}[h!]
    \centering
    \includegraphics[scale=0.7]{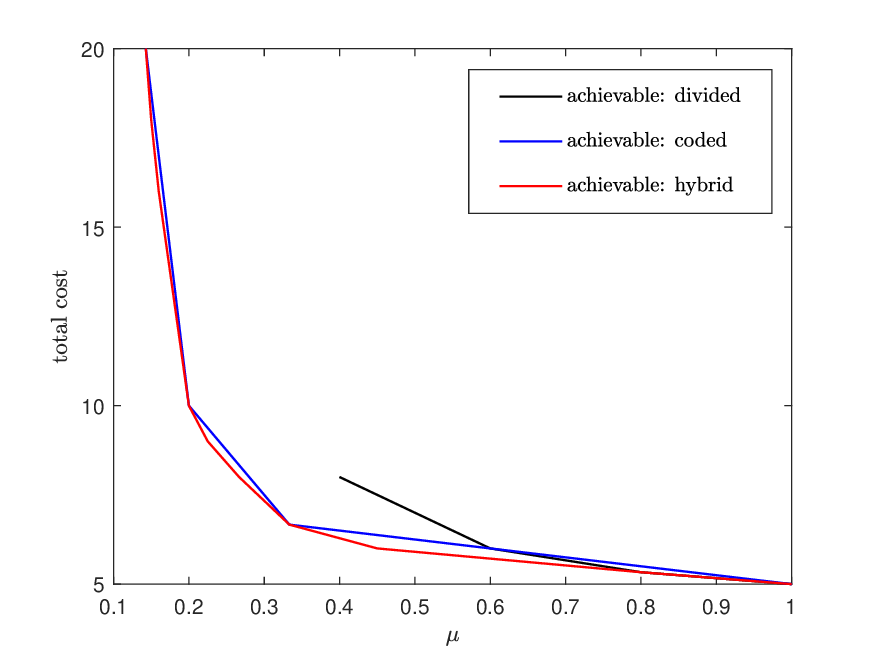}
    \caption{Lowest achievable costs of coded, divided and hybrid schemes for $N=10$.}
    \label{fig4}
    \vspace*{-0.4cm}
\end{figure}

The set of basic achievable $\left(\mu=\frac{r}{NK_r},C_T(\mu)\right)$ pairs on the lower convex hull corresponds to storage constraints $\mu$ with $(R,K_R)$ pairs corresponding to $R=N=10$, $R=N-1=9$ and $R=N-2=8$ with $K_R$ values that satisfy $(R-K_R-1)\mod2=0$, as marked in green in Fig.~\ref{fig2}. Note that the minimum achievable costs of divided and coded schemes are determined by the lower convex hull of the points marked in blue and red in Fig.~\ref{fig2}, respectively, which are subsets of all points considered in the lower convex hull search of the hybrid scheme, which clearly results in lower achievable costs as shown in Fig.~\ref{fig4}.

\subsection{Example}

In this section, we describe how the PRUW process is carried out in an arbitrary setting with given $N$ and $\mu$. Consider an example with $N=8$ databases and $\mu=0.7$. The first step is to find the basic achievable $\left(\mu=\frac{R}{NK_R},C_T(\mu)\right)$ pairs of $N=8$ that lie on the lower convex hull boundary. Fig.~\ref{fig7}(a) shows the $(R,K_R)$ pairs and the corresponding $\mu$s of such pairs. The required storage constraint $\mu=0.7$ is in between $0.44$ and $0.75$, which correspond to $(R,K_R)$ pairs $(7,2)$ and $(6,1)$, respectively. Therefore, the PRUW scheme for $N=8$, $\mu=0.7$ is obtained by the following steps:

\begin{figure}
\centering
\begin{subfigure}[b]{0.3\columnwidth}
\includegraphics[width=\columnwidth]{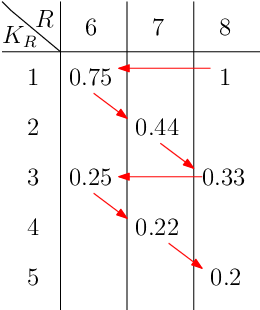}
\vspace{0.4cm}
\caption{$\mu=\frac{r}{NK_r}$ on the boundary.}
\end{subfigure}
\begin{subfigure}[b]{0.65\columnwidth}
\includegraphics[width=\columnwidth]{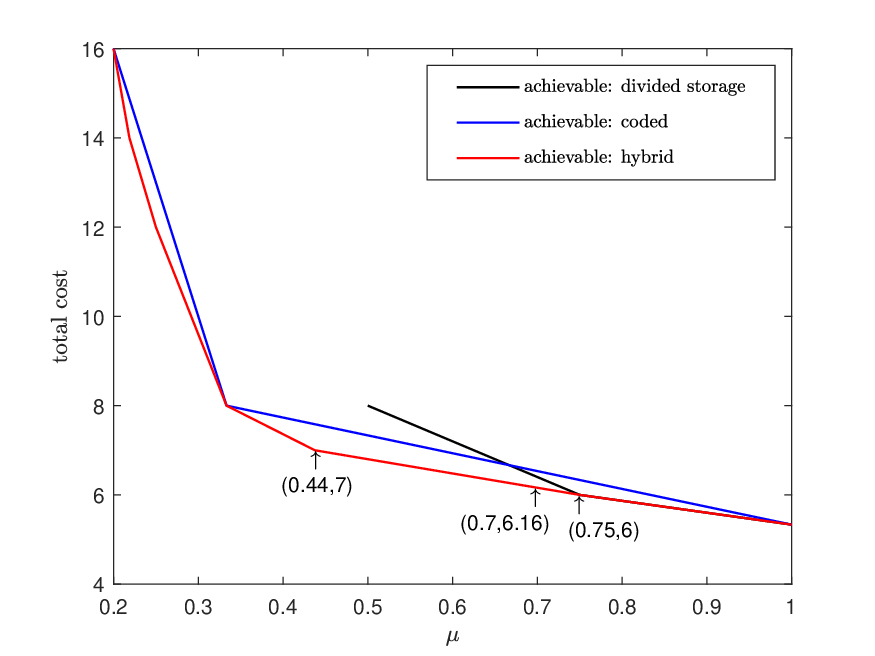}
\caption{Achievable total cost at $\mu=0.7$.}
\end{subfigure}
\caption{Example with $N=8$.}
\label{fig7}
\vspace*{-0.4cm}
\end{figure}

\begin{enumerate}
    \item $\gamma L$ bits of all submodels are stored according to the proposed storage mechanism corresponding to $(R,K_R)=(7,2)$, and the rest of the $(1-\gamma)L$ bits of all submodels are stored according to $(R,K_R)=(6,1)$. Therefore, $\gamma L$ bits of the required submodel are updated using the scheme corresponding to $(R,K_R)=(7,2)$, and the rest of the bits are updated by the scheme corresponding to $(6,1)$. In order to find the value of $\gamma$, we equate the total storage of each database to the given constraint, i.e.,
    \begin{align}
        \gamma ML\times\frac{7}{8}\times\frac{1}{2}+(1-\gamma)ML\times\frac{6}{8}=0.7ML
    \end{align}
    which gives $\gamma=0.16$.
    \item Let $L_1=0.16L$ and $L_2=0.84L$. $L_1$ bits of each submodel is divided into $8$ sections and labeled $1,\dotsc,8$. Sections $n:(n+6)\!\!\mod8$ are allocated to database $n$ for $n\in\{1,\dotsc,N\}$. Each database uses the storage in \eqref{storage} with $K=2$ and $y=x=\frac{R-K_R-1}{2}=2$ to store each subpacket of all sections allocated to it. Then, the PRUW scheme described in Section~\ref{pruw-general} is applied to read/write to the $L_1$ bits of the required submodel.
     \item The same process is carried out on the rest of the $L_2$ bits with the scheme corresponding to $(6,1)$.
\end{enumerate}

The total costs incurred by the two schemes are $C_{T_1}=\frac{4R}{R-K_R-1}=\frac{4\times 7}{7-2-1}=7$ and $C_{T_2}=\frac{4R}{R-K_R-1}=\frac{4\times 6}{6-1-1}=6$, respectively. Therefore, the total cost of $N=8$ and $\mu=0.7$ is $C_T=\frac{\gamma LC_{T_1}+(1-\gamma)LC_{T_2}}{L}=6.16$,
which is shown in Fig.~\ref{fig7}(b).

\section{Discussion and Conclusions}

In this work, we considered the problem of information-theoretically private FSL with storage constrained databases. We considered both heterogeneous and homogeneous storage constraints, and proposed schemes to perform private FSL in both settings, with the goal of minimizing the total communication cost while guaranteeing the privacy of the updating submodel index and the values of the updates. As the main result, we proposed a PRUW scheme and a storage mechanism that is applicable to any given set of heterogeneous storage constraints, and a different storage mechanism built upon the same PRUW scheme for homogeneous storage constraints. Although the general scheme for heterogeneous constraints is not applicable to homogeneous constraints in general, the range of costs achieved by the latter are lower than what is achieved by the former. This is because the scheme proposed for homogeneous constraints is able to eliminate an inefficient MDS coding structure in its storage, which is not possible in the heterogeneous case if the scheme must serve any given set of heterogeneous storage constraints.

\appendix

\section{Proof of Lemma~\ref{lem0}}\label{pflem1}

When $\alpha=1$, the only coding parameter used in the entire storage is $\lfloor k\rfloor$, which modifies the maximum number of replications in \eqref{repl} as,
\begin{align}
    R\leq\frac{\sum_{n=1}^N \mu(n)ML}{\frac{ML}{\lfloor k\rfloor}}=\lfloor k\rfloor p=s.
\end{align}
If $s\in\mathbb{Z}^+$, all parameters are encoded with the $(\lfloor k\rfloor,s)$ MDS code. However, if $s\notin\mathbb{Z}^+$ we need to find the optimum fractions $\beta$ and storage allocations\footnote{$\hat{\mu}_1(n)$ and $\hat{\mu}_2(n)$ are the storage allocations for $(\lfloor k\rfloor,\lfloor s\rfloor)$ and $(\lfloor k\rfloor,\lceil s\rceil)$ MDS codes in database $n$.} $\hat{\mu}_1(n)$, $\hat{\mu}_2(n)$ of each submodel to be encoded with $(\lfloor k\rfloor,\lfloor s\rfloor)$ and $(\lfloor k\rfloor,\lceil s\rceil)$ MDS codes that minimize the modified total cost in \eqref{min} given by,
\begin{align}\label{minmod}
    C=\beta C_T(\lfloor k\rfloor,\lfloor s\rfloor)+(1-\beta) C_T(\lfloor k\rfloor,\lceil s\rceil).
\end{align}
The total storage allocations in the entire system of databases for the two MDS codes must satisfy (modified \eqref{eq1} and \eqref{eq2}),
\begin{align}
    \sum_{n=1}^N \hat{\mu}_1(n)&=\frac{\beta}{\lfloor k\rfloor}\lfloor s\rfloor\label{sp1}\\
    \sum_{n=1}^N \hat{\mu}_2(n)&=\frac{1-\beta}{\lfloor k\rfloor}\lceil s\rceil.\label{sp2}
\end{align}
Moreover, since
\begin{align}
    \sum_{n=1}^N \hat{\mu}_1(n)+\sum_{n=1}^N \hat{\mu}_2(n)=\sum_{n=1}^N \mu(n)=p=\frac{\lceil s\rceil-\beta}{\lfloor k\rfloor},
\end{align}
$\beta$ is fixed at $\beta=\lceil s\rceil-s$, and the resulting total cost is given by \eqref{c1}. Note that the total cost in \eqref{minmod} does not depend on each individual $\hat{\mu}_1(n)$ and $\hat{\mu}_2(n)$. Therefore, any set of $\hat{\mu}_1(n)$ and $\hat{\mu}_2(n)$ satisfying \eqref{sp1} and \eqref{sp2} with $\beta=\lceil s\rceil-s$ along with the following constraints (modified \eqref{neq1} and \eqref{neq2}) results in the same total cost. Modified \eqref{neq1} and \eqref{neq2} are given by,
\begin{align}
    \hat{\mu}_1(n)&\leq \frac{\lceil s\rceil-s}{\lfloor k\rfloor}\label{sp3}\\
    \hat{\mu}_2(n)&\leq \frac{s-\lfloor s\rfloor}{\lfloor k\rfloor}.\label{sp4}
\end{align}
Therefore, it remains to prove that $\hat{\mu}_1(n)$ and $\hat{\mu}_1(n)$ for each $n$ provided in Lemma~\ref{lem0} satisfy \eqref{sp1}-\eqref{sp4} when $\beta=\lceil s\rceil-s$. Note that \eqref{mu11} results in,
\begin{align}
    \sum_{n=1}^N \hat{\mu}_1(n)=\sum_{n=1}^N \Tilde{m}(n)+\Tilde{\gamma}(p-\sum_{n=1}^N \Tilde{m}(n)-\sum_{n=1}^N \Tilde{h}(n))=\frac{\lfloor s\rfloor}{\lfloor k\rfloor}(\lceil s\rceil-s),
\end{align}
which proves \eqref{sp1}. Similarly, summing up \eqref{mu21} results in \eqref{sp2}. Assuming that $0\leq\Tilde{\gamma}\leq1$, \eqref{mu11} gives,
\begin{align}
    \hat{\mu}_1(n)&\leq \mu(n)-\Tilde{h}(n)\\
    &\leq\begin{cases}
        \frac{\lceil s\rceil-s}{\lfloor k\rfloor}, & \text{if $\mu(n)\geq \frac{\lceil s\rceil-s}{\lfloor k\rfloor}$}\\
        \mu(n), & \text{if $\mu(n)< \frac{\lceil s\rceil-s}{\lfloor k\rfloor}$},
    \end{cases}
\end{align}
which proves \eqref{sp3}. A similar proof is valid for \eqref{sp4} as well, given that $0\leq\Tilde{\gamma}\leq1$, which is proved next. Let $\mathcal{N}$ and $\mathcal{D}$ denote the numerator and the denominator of $\Tilde{\gamma}$ in \eqref{tildegamma}. Note that $\Tilde{m}(n)+\Tilde{h}(n)$ is given by,
\begin{align}
    \Tilde{m}(n)+\Tilde{h}(n)=\begin{cases}
        \mu(n)-\frac{s-\lfloor s\rfloor}{\lfloor k\rfloor}, & \text{if $\frac{s-\lfloor s\rfloor}{\lfloor k\rfloor}\leq \mu(n)\leq\frac{\lceil s\rceil-s}{\lfloor k\rfloor}$}\\
        \mu(n)-\frac{\lceil s\rceil-s}{\lfloor k\rfloor}, & \text{if $\frac{\lceil s\rceil-s}{\lfloor k\rfloor}\leq \mu(n)\leq\frac{s-\lfloor s\rfloor}{\lfloor k\rfloor}$}\\
        2\mu(n)-\frac{1}{\lfloor k\rfloor}, & \text{if $\frac{\lceil s\rceil-s}{\lfloor k\rfloor}, \frac{s-\lfloor s\rfloor}{\lfloor k\rfloor}\leq \mu(n)$}\\
        0, & \text{if $\frac{\lceil s\rceil-s}{\lfloor k\rfloor}, \frac{s-\lfloor s\rfloor}{\lfloor k\rfloor}\geq \mu(n)$}
    \end{cases}\leq\mu(n),
\end{align}
since $\max_n \mu(n)\leq\frac{1}{k}\leq\frac{1}{\lfloor k\rfloor}$, and the equality holds only if $\mu(n)=\frac{1}{\lfloor k\rfloor}$. Therefore,
\begin{align}
    \sum_{n=1}^N \Tilde{m}(n)+\sum_{n=1}^N \Tilde{h}(n)\leq p,
\end{align}
and the equality holds only if $\mu(n)=\frac{1}{\lfloor k\rfloor}$, $\forall n$, which specifies a set of homogeneous storage constraints. Since we only consider heterogeneous storage constraints, we have, $\sum_{n=1}^N \Tilde{m}(n)+\sum_{n=1}^N \Tilde{h}(n)< p$, which proves $\mathcal{D}>0$. Therefore, it remains to prove that $\mathcal{N}\leq\mathcal{D}$ and $\mathcal{N}\geq0$, to prove $0\leq\Tilde{\gamma}\leq1$. Note that $\mathcal{N}\leq\mathcal{D}$ is equivalent to $\frac{\lfloor s\rfloor}{\lfloor k\rfloor}(\lceil s\rceil-s)\leq p-\sum_{n=1}^N \Tilde{h}(n)$. Let $V$ be defined as,
\begin{align}
    V=\sum_{n=1}^N \mathbf{1}_{\{\mu(n)\geq \frac{\lceil s\rceil-s}{\lfloor k\rfloor}\}}.
\end{align}
where $\mathbf{1}_{\{\cdot\}}$ is the indicator function. From the definition of $\Tilde{h}(n)$ in \eqref{tildemh} and since $\mu(n)\leq\frac{1}{\lfloor k\rfloor}$,
\begin{align}
    \sum_{n=1}^N\Tilde{h}(n)\leq \frac{V}{\lfloor k\rfloor}-\frac{V(\lceil s\rceil-s)}{\lfloor k\rfloor}.\label{in1}
\end{align}
Moreover, since the total available space in the $V$ databases must not exceed $p$,
\begin{align}
    \sum_{n=1}^N\Tilde{h}(n)+\frac{V(\lceil s\rceil-s)}{\lfloor k\rfloor}\leq p\quad \implies\quad \sum_{n=1}^N\Tilde{h}(n)\leq p-\frac{V(\lceil s\rceil-s)}{\lfloor k\rfloor}\label{in2}
\end{align}
must be satisfied. Note that the upper bound on $\sum_{n=1}^N \Tilde{h}(n)$ in \eqref{in1} is tighter than that of \eqref{in2} when $V\leq \lfloor k\rfloor p=s$ and vice versa. Therefore, the highest upper bound on $\sum_{n=1}^N \Tilde{h}(n)$ is given by,
\begin{align}
    \sum_{n=1}^N \Tilde{h}(n)\leq\begin{cases}
        \frac{\lfloor s\rfloor}{\lfloor k\rfloor}(s-\lfloor s\rfloor), & \text{if $V<s$}\\
        p-\frac{\lceil s\rceil}{\lfloor k\rfloor}(\lceil s\rceil-s), & \text{if $V>s$}
    \end{cases},
\end{align}
since $V\in\mathbb{Z}^+$ and $s\notin\mathbb{Z}^+$,\footnote{If $s\in\mathbb{Z}^+$ all parameters in all submodels will be $(\lfloor k\rfloor,s)$ MDS coded, which does not require any fractions/storage allocations to be calculated.} which proves,
\begin{align}
    p-\sum_{n=1}^N \Tilde{h}(n)&\geq\begin{cases}
        p-\frac{\lfloor s\rfloor}{\lfloor k\rfloor}(s-\lfloor s\rfloor), & \text{if $V<s$}\\
        \frac{\lceil s\rceil}{\lfloor k\rfloor}(\lceil s\rceil-s), & \text{if $V>s$}
    \end{cases}\\
    &\geq\frac{\lfloor s\rfloor}{\lfloor k\rfloor}(\lceil s\rceil-s)\quad \implies\quad \mathcal{N}\leq\mathcal{D}.
\end{align}
To prove $\mathcal{N}\geq0$, we need to show that $\sum_{n=1}^N \Tilde{m}(n)\leq\frac{\lfloor s\rfloor}{\lfloor k\rfloor}(\lceil s\rceil-s)$. Let $Y$ be defined as,
\begin{align}
    Y=\sum_{n=1}^N \mathbf{1}_{\{\mu(n)\geq \frac{s-\lfloor s\rfloor}{\lfloor k\rfloor}\}}.
\end{align}
Then, similar to \eqref{in1} and \eqref{in2}, we have,
\begin{align}
    \sum_{n=1}^N\Tilde{m}(n)&\leq \frac{Y}{\lfloor k\rfloor}-\frac{Y(s-\lfloor s\rfloor)}{\lfloor k\rfloor}\label{in1m}\\
    \sum_{n=1}^N\Tilde{m}(n)&\leq p-\frac{Y(s-\lfloor s\rfloor)}{\lfloor k\rfloor}\label{in2m},
\end{align}
and \eqref{in1m} provides a tighter upper bound on $\sum_{n=1}^N \Tilde{m}(n)$ compared to \eqref{in2m} when $Y\leq \lfloor k\rfloor p=s$, and vice versa. Therefore, the highest upper bound on $\sum_{n=1}^N \Tilde{m}(n)$ is given by,
\begin{align}
    \sum_{n=1}^N \Tilde{m}(n)&\leq\begin{cases}
        \frac{\lfloor s\rfloor}{\lfloor k\rfloor}(\lceil s\rceil-s), & \text{if $Y<s$}\\
        p-\frac{\lceil s\rceil}{\lfloor k\rfloor}(s-\lfloor s\rfloor), & \text{if $Y>s$}
    \end{cases} \\
    &=\frac{\lfloor s\rfloor}{\lfloor k\rfloor}(\lceil s\rceil-s)\quad \implies\quad \mathcal{N}\geq0
\end{align}
since $Y\in\mathbb{Z}^+$ and $s\notin\mathbb{Z}^+$, which completes the proof of $0\leq\Tilde{\gamma}\leq1$.

\section{Proof of Lemma~\ref{lem1}}\label{pflem2}

Here, we prove that the storage allocations provided in Lemma~\ref{lem1} satisfy \eqref{eq1}-\eqref{sum} for the case where $\alpha<1$. The storage allocations $\hat{\mu}_1(n)$, $\hat{\mu}_2(n)$, $\bar{\mu}_1(n)$ and $\bar{\mu}_2(n)$ correspond to MDS codes $(\lfloor k\rfloor,\lfloor r\rfloor)$, $(\lfloor k\rfloor,\lceil r\rceil)$, $(\lceil k\rceil,\lfloor r\rfloor)$ and $(\lceil k\rceil,\lceil r\rceil)$, respectively. We first determine the storage allocations corresponding to the two coding parameters $\lfloor k\rfloor$ and $\lceil k\rceil$, i.e., the total storage allocations for the two pairs of MDS codes $\{(\lfloor k\rfloor,\lfloor r\rfloor),(\lfloor k\rfloor,\lceil r\rceil)\}$ and $\{(\lceil k\rceil,\lfloor r\rfloor),(\lceil k\rceil,\lceil r\rceil)\}$ given by,
\begin{align}
    \hat{\mu}(n)&=\hat{\mu}_1(n)+\hat{\mu}_2(n),\quad n\in\{1,\dotsc,N\}\\
    \bar{\mu}(n)&=\bar{\mu}_1(n)+\bar{\mu}_2(n),\quad n\in\{1,\dotsc,N\}.
\end{align}
Then, the constraints \eqref{eq1}-\eqref{sum} impose the following constraints on $\hat{\mu}(n)$ and $\bar{\mu}(n)$,
\begin{align}
    \sum_{n=1}^N \hat{\mu}(n)&=\frac{\alpha}{\lfloor k\rfloor}(\lceil r\rceil-\beta)\label{comb1}\\
    \sum_{n=1}^N \bar{\mu}(n)&=\frac{1-\alpha}{\lceil k\rceil}(\lceil r\rceil-\delta)\label{comb2}\\
    \hat{\mu}(n)&\leq \frac{\alpha}{\lfloor k\rfloor},\quad n\in\{1,\dotsc,N\} \label{comb3}\\
    \bar{\mu}(n)&\leq \frac{1-\alpha}{\lceil k\rceil},\quad n\in\{1,\dotsc,N\} \label{comb4}\\
    \hat{\mu}(n)+\Bar{\mu}(n)&=\mu(n),\quad \forall n.\label{comb5}
\end{align}
The combination of \eqref{comb1}-\eqref{comb5} gives,
\begin{align}
    \mu(n)-\frac{1-\alpha}{\lceil k\rceil}&\leq\hat{\mu}(n)\leq\frac{\alpha}{\lfloor k\rfloor},\quad n\in\{1,\dotsc,N\}\label{combb1}\\
    \mu(n)-\frac{\alpha}{\lfloor k\rfloor}&\leq\bar{\mu}(n)\leq\frac{1-\alpha}{\lceil k\rceil},\quad n\in\{1,\dotsc,N\}\label{combb2}\\
    \frac{\alpha}{\lfloor k\rfloor}(\lceil r\rceil-\beta)&+\frac{1-\alpha}{\lceil k\rceil}(\lceil r\rceil-\delta)=p.\label{combb3}
\end{align}
Based on \eqref{combb1} and \eqref{combb2}, for each $n\in\{1,\dotsc,N\}$ define,
\begin{align}\label{mh1}
    m(n)=\left[\mu(n)-\frac{1-\alpha}{\lceil k\rceil}\right]^+,\quad h(n)=\left[\mu(n)-\frac{\alpha}{\lfloor k\rfloor}\right]^+.
\end{align}
Then, the total storage allocations corresponding to the two coding parameters $\lfloor k\rfloor$ and $\lceil k\rceil$ in database $n$, $n\in\{1,\dotsc,N\}$, are chosen as, 
\begin{align}
    \hat{\mu}(n)&=m(n)+(\mu(n)-m(n)-h(n))\gamma\label{muhat11}\\
    \bar{\mu}(n)&=h(n)+(\mu(n)-m(n)-h(n))(1-\gamma)\label{mubar11},
\end{align}
where
\begin{align}
    \gamma=\frac{\frac{\alpha}{\lfloor k\rfloor}(\lceil r\rceil-\beta)-\sum_{n=1}^N m(n)}{p-\sum_{n=1}^N m(n)-\sum_{n=1}^N h(n)}.\label{gama1}
\end{align}

\textbf{Claim~1:} For each $n\in\{1,\dotsc,N\}$, $\hat{\mu}(n)$ and $\bar{\mu}(n)$, in \eqref{muhat11} and \eqref{mubar11} (same as \eqref{ee1} and \eqref{eee1}) satisfy \eqref{comb1}-\eqref{comb5} for those $\alpha,\beta,\delta$ stated in Lemma~\ref{lem1}.

\begin{Proof}
Summing each $\hat{\mu}(n)$ term in \eqref{muhat11} yields,
\begin{align}
    \sum_{n=1}^N \hat{\mu}(n)&=\sum_{n=1}^N m(n)+\gamma(p-\sum_{n=1}^N m(n)-\sum_{n=1}^N h(n))=\frac{\alpha}{\lfloor k\rfloor}(\lceil r\rceil-\beta),
\end{align}
from \eqref{gama1}, which proves \eqref{comb1}. A similar proof results in \eqref{comb2}. Assuming that $0\leq\gamma\leq1$, \eqref{muhat11} can be upper bounded by,
\begin{align}
    \hat{\mu}(n)&\leq\mu(n)-h(n)\\ 
    &=\begin{cases}
        \frac{\alpha}{\lfloor k\rfloor}, & \text{if $\mu(n)\geq\frac{\alpha}{\lfloor k\rfloor}$}\\
        \mu(n), & \text{if $\mu(n)<\frac{\alpha}{\lfloor k\rfloor}$}
    \end{cases}\\
    &\leq\frac{\alpha}{\lfloor k\rfloor},
\end{align}
which proves \eqref{comb3}. Similarly, \eqref{comb4} is proven by considering $\Bar{\mu}(n)$. \eqref{comb5} is obvious from \eqref{muhat11} and \eqref{mubar11}. Hence, it remains to prove that $0\leq\gamma\leq1$. Similar to the proof of $0\leq\Tilde{\gamma}\leq1$ in Lemma~\ref{lem0}, let $\mathcal{N}$ and $\mathcal{D}$ denote the numerator and the denominator of $\gamma$ in \eqref{gama1}. Then, $m(n)+h(n)$ is given by,
\begin{align}
    m(n)+h(n)&=\begin{cases}
        \mu(n)-\frac{\alpha}{\lfloor k\rfloor}, & \text{if $\frac{\alpha}{\lfloor k\rfloor}\leq \mu(n)\leq\frac{1-\alpha}{\lceil k\rceil}$}\\
        \mu(n)-\frac{1-\alpha}{\lceil k\rceil}, & \text{if $\frac{1-\alpha}{\lceil k\rceil}\leq \mu(n)\leq\frac{\alpha}{\lfloor k\rfloor}$}\\
        2\mu(n)-\frac{\alpha}{\lfloor k\rfloor}-\frac{1-\alpha}{\lceil k\rceil}, & \text{if $\frac{\alpha}{\lfloor k\rfloor}, \frac{1-\alpha}{\lceil k\rceil}\leq \mu(n)$}\\
        0, & \text{if $\frac{\alpha}{\lfloor k\rfloor}, \frac{1-\alpha}{\lceil k\rceil}\geq \mu(n)$}
    \end{cases}\\
    &\leq\mu(n),
\end{align}
if $\mu(n)\leq\frac{\alpha}{\lfloor k\rfloor}+\frac{1-\alpha}{\lceil k\rceil}$, $\forall n$, i.e., $\alpha\geq\frac{\lfloor k\rfloor}{k}(\lceil k\rceil-k)$, which is the constraint on $\alpha$ stated in Lemma~\ref{lem1}. Therefore,
\begin{align}\label{deno}
    \sum_{n=1}^N m(n)+\sum_{n=1}^N h(n)\leq p,
\end{align}
and the equality holds only if $\mu(n)=\frac{1}{k}$, $\forall n$ and $\alpha=\frac{\lfloor k\rfloor}{k}(\lceil k\rceil-k)$, which specifies a set of homogeneous storage constraints. Since we only consider heterogeneous storage constraints, \eqref{deno} is satisfied with strict inequality when $\alpha$ is chosen such that $\frac{\lfloor k\rfloor}{k}(\lceil k\rceil-k)\leq\alpha<1$, which proves $\mathcal{D}>0$. Therefore, it remains to prove that $\mathcal{N}\leq\mathcal{D}$ and $\mathcal{N}\geq0$, to prove $0\leq\gamma\leq1$. Note that $\mathcal{N}\leq\mathcal{D}$ is equivalent to $\frac{\alpha}{\lfloor k\rfloor}(\lceil r\rceil-\beta)\leq p-\sum_{n=1}^N h(n)$. Let $V$ be defined as,
\begin{align}
    V=\sum_{n=1}^N \mathbf{1}_{\{\mu(n)\geq \frac{\alpha}{\lfloor k\rfloor}\}}.
\end{align}
Therefore from the definition of $h(n)$ in \eqref{mh1} and since $\mu(n)\leq\frac{1}{k}$, $\forall n$,
\begin{align}
    \sum_{n=1}^N h(n)\leq \frac{V}{k}-\frac{V\alpha}{\lfloor k\rfloor}.\label{in1no}
\end{align}
Moreover, since the total available space in the $V$ databases must not exceed $p$,
\begin{align}
    \sum_{n=1}^N h(n)+\frac{V\alpha}{\lfloor k\rfloor}\leq p\quad \implies\quad \sum_{n=1}^N h(n)\leq p-\frac{V\alpha}{\lfloor k\rfloor}\label{in2no}
\end{align}
must be satisfied. The upper bound on $\sum_{n=1}^N h(n)$ in \eqref{in1no} is tighter than that of \eqref{in2no} when $V\leq kp=r$ and vice versa. Thus, the highest upper bound on $\sum_{n=1}^N 
 h(n)$ is given by,
\begin{align}
    \sum_{n=1}^N h(n)\leq\begin{cases}
        \frac{\lfloor r\rfloor}{k}-\frac{\lfloor r\rfloor\alpha}{\lfloor k\rfloor}, & \text{if $V<r$}\\
        p-\frac{\lceil r\rceil\alpha}{\lfloor k\rfloor}, & \text{if $V>r$}
    \end{cases},
\end{align}
since $V\in\mathbb{Z}^+$, which gives the tightest upper bound on $p-\sum_{n=1}^N h(n)$ for a general set of $\{\mu(n)\}_{n=1}^N$ as,
\begin{align}
    p-\sum_{n=1}^N h(n)&\geq\begin{cases}
        p-\frac{\lfloor r\rfloor}{k}+\frac{\lfloor r\rfloor\alpha}{\lfloor k\rfloor}, & \text{if $V<r$}\\
        \frac{\lceil r\rceil\alpha}{\lfloor k\rfloor}, & \text{if $V>r$}
    \end{cases}.
\end{align}
Therefore, in order to satisfy $\mathcal{N}\leq\mathcal{D}$ for any given set of $\{\mu(n)\}_{n=1}^N$, i.e.,
\begin{align}
    p-\sum_{n=1}^N h(n)&\geq\begin{cases}
        p-\frac{\lfloor r\rfloor}{k}+\frac{\lfloor r\rfloor\alpha}{\lfloor k\rfloor}, & \text{if $V<r$}\\
        \frac{\lceil r\rceil\alpha}{\lfloor k\rfloor}, & \text{if $V>r$}
    \end{cases}\\
    &\geq\frac{\alpha}{\lfloor k\rfloor}(\lceil r\rceil-\beta)
\end{align}
it requires $\beta$ to satisfy $\beta\geq1-\frac{\lfloor k\rfloor}{k\alpha}(r-\lfloor r\rfloor)$, which is the constraint given on $\beta$ in Lemma~\ref{lem1}. To prove $\mathcal{N}\geq0$, we need to show that $\sum_{n=1}^N m(n)\leq\frac{\alpha}{\lfloor k\rfloor}(\lceil r\rceil-\beta)$. Let $Y$ be defined as,
\begin{align}
    Y=\sum_{n=1}^N \mathbf{1}_{\{\mu(n)\geq \frac{1-\alpha}{\lceil k\rceil}\}}.
\end{align}
Then, similar to \eqref{in1no} and \eqref{in2no}, we have,
\begin{align}
    \sum_{n=1}^N m(n)&\leq \frac{Y}{k}-\frac{Y(1-\alpha)}{\lceil k\rceil}\label{in1mno}\\
    \sum_{n=1}^N m(n)&\leq p-\frac{Y(1-\alpha)}{\lceil k\rceil}\label{in2mno},
\end{align}
and \eqref{in1mno} provides a tighter upper bound on $\sum_{n=1}^N m(n)$ compared to \eqref{in2mno} when $Y\leq kp=r$, and vice versa. Therefore, the highest upper bound on $\sum_{n=1}^N m(n)$ considering any arbitrary set of $\{\mu(n)\}_{n=1}^N$ is given by,
\begin{align}
    \sum_{n=1}^N m(n)&\leq\begin{cases}
        \frac{\lfloor r\rfloor}{k}-\frac{\lfloor r\rfloor(1-\alpha)}{\lceil k\rceil}, & \text{if $Y<r$}\\
        p-\frac{\lceil r\rceil(1-\alpha)}{\lceil k\rceil}, & \text{if $Y>r$}
    \end{cases} 
\end{align}
since $Y\in\mathbb{Z}^+$. Therefore, $\mathcal{N}\geq0$ is satisfied for any set of $\{\mu(n)\}_{n=1}^N$ if
\begin{align}
    \sum_{n=1}^N m(n)&\leq\begin{cases}
        \frac{\lfloor r\rfloor}{k}-\frac{\lfloor r\rfloor(1-\alpha)}{\lceil k\rceil}, & \text{if $Y<r$}\\
        p-\frac{\lceil r\rceil(1-\alpha)}{\lceil k\rceil}, & \text{if $Y>r$}
        \end{cases}\\
        &\leq\frac{\alpha}{\lfloor k\rfloor}(\lceil r\rceil-\beta)
\end{align}
is satisfied, which requires $\delta\geq1-\frac{\lceil k\rceil}{k(1-\alpha)}(r-\lfloor r\rfloor)$. This is the constraint stated in Lemma~\ref{lem1}, which is derived using \eqref{combb3}. Therefore, $0\leq\gamma\leq1$ is satisfied by any given set of arbitrary heterogeneous storage constraints $\{\mu(n)\}_{n=1}^N$ when
\begin{align}
    1>\alpha&\geq\frac{\lfloor k\rfloor}{k}(\lceil k\rceil-k)\label{alpha}\\
    1\geq\beta&\geq\left[1-\frac{\lfloor k\rfloor}{k\alpha}(r-\lfloor r\rfloor)\right]^+\label{beta}\\
    1\geq\delta&\geq\left[1-\frac{\lceil k\rceil}{k(1-\alpha)}(r-\lfloor r\rfloor)\right]^+.\label{delta}
\end{align}
which completes the proof of Claim~1.
\end{Proof}

The above proof finalizes the storage allocations corresponding to coding parameters $\lfloor k\rfloor$ and $\lceil k\rceil$. Next, we find the storage allocations corresponding to each of the two MDS codes relevant to each coding parameter, i.e., $(\lfloor k\rfloor,\lfloor r\rfloor)$ and $(\lfloor k\rfloor,\lceil r\rceil)$ corresponding to $\lfloor k\rfloor$ and $(\lceil k\rceil,\lfloor r\rfloor)$ and $(\lceil k\rceil,\lceil r\rceil)$ corresponding to $\lceil k\rceil$. In other words, we further divide $\hat{\mu}(n)$ into $\hat{\mu}_1(n),\hat{\mu}_2(n)$ and $\bar{\mu}(n)$ into $\bar{\mu}_1(n),\bar{\mu}_2(n)$ such that $\hat{\mu}_1(n),\hat{\mu}_2(n),\bar{\mu}_1(n),\bar{\mu}_2(n)$ satisfy \eqref{eq1}-\eqref{sum}. Note that the constraints \eqref{neq1}-\eqref{neq4} result in,
\begin{align}
\hat{\mu}(n)-\frac{\alpha(1-\beta)}{\lfloor k\rfloor}&\leq\hat{\mu}_1(n)\leq\frac{\alpha\beta}{\lfloor k\rfloor}\label{hatcomb1}\\
\hat{\mu}(n)-\frac{\alpha\beta}{\lfloor k\rfloor}&\leq\hat{\mu}_2(n)\leq\frac{\alpha(1-\beta)}{\lfloor k\rfloor}\label{hatcomb2}\\
    \bar{\mu}(n)-\frac{(1-\alpha)(1-\delta)}{\lceil k\rceil}&\leq\bar{\mu}_1(n)\leq\frac{(1-\alpha)\delta}{\lceil k\rceil}\label{hatcomb3}\\
\bar{\mu}(n)-\frac{(1-\alpha)\delta}{\lceil k\rceil}&\leq\bar{\mu}_2(n)\leq\frac{(1-\alpha)(1-\delta)}{\lceil k\rceil},\label{hatcomb4}
\end{align}
for each $n\in\{1,\dotsc,N\}$. Based on \eqref{hatcomb1}-\eqref{hatcomb4}, for each $n\in\{1,\dotsc,N\}$ define
\begin{align}
    \hat{m}(n)&=\left[\hat{\mu}(n)-\frac{\alpha(1-\beta)}{\lfloor k\rfloor}\right]^+,\ \quad\qquad \hat{h}(n)=\left[\hat{\mu}(n)-\frac{\alpha\beta}{\lfloor k\rfloor}\right]^+\label{hatmh}\\
    \bar{m}(n)&=\left[\bar{\mu}(n)-\frac{(1-\alpha)(1-\delta)}{\lceil k\rceil}\right]^+,\quad \bar{h}(n)=\left[\bar{\mu}(n)-\frac{(1-\alpha)\delta}{\lceil k\rceil}\right]^+.\label{barmh}
\end{align}
Then, define the storage allocations in database $n$, $n\in\{1,\dotsc,N\}$ corresponding to MDS codes $(\lfloor k\rfloor,\lfloor r\rfloor)$, $(\lfloor k\rfloor,\lceil r\rceil)$, $(\lceil k\rceil,\lfloor r\rfloor)$ and $(\lceil k\rceil,\lceil r\rceil)$ as
\begin{align}
    \hat{\mu}_1(n)&=\begin{cases}
    \hat{\mu}(n)\beta, & \text{if $\beta\in\{0,1\}$}\\
    \hat{m}(n)+(\hat{\mu}(n)-\hat{m}(n)-\hat{h}(n))\hat{\gamma}, & \text{if $\beta\in(0,1)$}
    \end{cases}\label{a1}\\
    \hat{\mu}_2(n)&=\begin{cases}
    \hat{\mu}(n)(1-\beta), & \text{if $\beta\in\{0,1\}$}\\
    \hat{h}(n)+(\hat{\mu}(n)-\hat{m}(n)-\hat{h}(n))(1-\hat{\gamma}), & \text{if $\beta\in(0,1)$}
    \end{cases}\label{a2}\\
    \bar{\mu}_1(n)&=\begin{cases}
    \bar{\mu}(n)\delta, & \text{if $\delta\in\{0,1\}$}\\
    \bar{m}(n)+(\bar{\mu}(n)-\bar{m}(n)-\bar{h}(n))\bar{\gamma}, & \text{if $\delta\in(0,1)$}
    \end{cases}\label{a3}\\
    \bar{\mu}_2(n)&=\begin{cases}
    \bar{\mu}(n)(1-\delta), & \text{if $\delta\in\{0,1\}$}\\
    \bar{h}(n)+(\bar{\mu}(n)-\bar{m}(n)-\bar{h}(n))(1-\bar{\gamma}), & \text{if $\delta\in(0,1)$}
    \end{cases},\label{a4}   
\end{align}
respectively, where,
\begin{align}
    \hat{\gamma}&=\frac{\frac{\alpha\beta}{\lfloor k\rfloor}\lfloor r\rfloor-\sum_{n=1}^N \hat{m}(n)}{\frac{\alpha}{\lfloor k\rfloor}(\lceil r\rceil-\beta)-\sum_{n=1}^N \hat{m}(n)-\sum_{n=1}^N \hat{h}(n)}\label{gamahat1}\\
    \bar{\gamma}&=\frac{\frac{(1-\alpha)\delta}{\lceil k\rceil}\lfloor r\rfloor-\sum_{n=1}^N \bar{m}(n)}{\frac{1-\alpha}{\lceil k\rceil}(\lceil r\rceil-\delta)-\sum_{n=1}^N \bar{m}(n)-\sum_{n=1}^N \bar{h}(n)}\label{gamabar1}.
\end{align}

\textbf{Claim~2:}  For each $n\in\{1,\dotsc,N\}$, $\hat{\mu}_1(n)$ and $\hat{\mu}_2(n)$, in \eqref{a1} and \eqref{a2} (same as \eqref{e1} and \eqref{muhat2}) satisfy \eqref{eq1}-\eqref{eq2} and \eqref{neq1}-\eqref{neq2} for those $\alpha,\beta,\delta$ stated in Lemma~\ref{lem1}.

\begin{Proof} \textit{Case~1: $\beta=1$}:
When $\beta=1$, all parameters that are $\lfloor k\rfloor$ coded are replicated in only $\lfloor r\rfloor$ databases since the fraction of submodels that are $(\lfloor k\rfloor,\lceil r\rceil)$ MDS coded is $\alpha(1-\beta)$ (see Table~\ref{codes}). Therefore, $\hat{\mu}_1(n)=\hat{\mu}(n)$ and $\hat{\mu}_2(n)=0$ for each $n\in\{1,\dotsc,N\}$. Then,
\begin{align}
    \sum_{n=1}^N\hat{\mu}_1(n)&=\sum_{n=1}^N\hat{\mu}(n)=\frac{\alpha}{\lfloor k\rfloor}\lfloor r\rfloor,
\end{align}
from \eqref{comb1}, which proves \eqref{eq1}. Moreover, $\hat{\mu}_2(n)=0$, $\forall n$ proves \eqref{eq2}. For each $n\in\{1,\dotsc,N\}$, \eqref{comb3} results in $\hat{\mu}_1(n)=\hat{\mu}(n)\leq\frac{\alpha}{\lfloor k\rfloor}$, which proves \eqref{neq1}, and $\hat{\mu}_2(n)=0$, $\forall n$ proves \eqref{neq2}.

\textit{Case~2: $\beta=0$}: The proof contains identical steps to the proof of Case~1.

\textit{Case~3: $\beta\in(0,1)$}: Summing each $\hat{\mu}_1(n)$ term in \eqref{a1} yields,
\begin{align}
    \sum_{n=1}^N \hat{\mu}_1(n)&=\sum_{n=1}^N \hat{m}(n)+\hat{\gamma}(\frac{\alpha}{\lfloor k\rfloor}(\lceil r\rceil-\beta)-\sum_{n=1}^N \hat{m}(n)-\sum_{n=1}^N \hat{h}(n))=\frac{\alpha\beta}{\lceil k\rceil}\lfloor r\rfloor,
\end{align}
from \eqref{gamahat1}, which proves \eqref{eq1}. A similar proof results in \eqref{eq2}. Assuming that $0\leq\gamma\leq1$, \eqref{a1} can be upper bounded by,
\begin{align}
    \hat{\mu}_1(n)&\leq\hat{\mu}(n)-\hat{h}(n)\\ 
    &=\begin{cases}
        \frac{\alpha\beta}{\lfloor k\rfloor}, & \text{if $\hat{\mu}(n)\geq\frac{\alpha\beta}{\lfloor k\rfloor}$}\\
        \hat{\mu}(n), & \text{if $\mu(n)<\frac{\alpha\beta}{\lfloor k\rfloor}$}
    \end{cases}\\
    &\leq\frac{\alpha\beta}{\lfloor k\rfloor},
\end{align}
which proves \eqref{neq1}. Similarly, \eqref{neq2} is proven by considering $\hat{\mu}_2(n)$. Hence, it remains to prove that $0\leq\hat{\gamma}\leq1$. Let $\mathcal{N}$ and $\mathcal{D}$ denote the numerator and the denominator of $\hat{\gamma}$ in \eqref{gamahat1}. Then, $\hat{m}(n)+\hat{h}(n)$ is given by,
\begin{align}
    \hat{m}(n)+\hat{h}(n)&=\begin{cases}
        \hat{\mu}(n)-\frac{\alpha(1-\beta)}{\lfloor k\rfloor}, & \text{if $\frac{\alpha(1-\beta)}{\lfloor k\rfloor}\leq \hat{\mu}(n)\leq\frac{\alpha\beta}{\lfloor k\rfloor}$}\\
        \hat{\mu}(n)-\frac{\alpha\beta}{\lfloor k\rfloor}, & \text{if $\frac{\alpha\beta}{\lfloor k\rfloor}\leq \hat{\mu}(n)\leq\frac{\alpha(1-\beta)}{\lfloor k\rfloor}$}\\
        2\hat{\mu}(n)-\frac{\alpha}{\lfloor k\rfloor}, & \text{if $\frac{\alpha\beta}{\lfloor k\rfloor}, \frac{\alpha(1-\beta)}{\lceil k\rceil}\leq \hat{\mu}(n)$}\\
        0, & \text{if $\frac{\alpha\beta}{\lfloor k\rfloor}, \frac{\alpha(1-\beta)}{\lfloor k\rfloor}\geq \hat{\mu}(n)$}
    \end{cases}\\
    &\leq\hat{\mu}(n),
\end{align}
since $\hat{\mu}(n)\leq\frac{\alpha}{\lfloor k\rfloor}$, $\forall n$, from \eqref{comb3}. Therefore, from \eqref{comb1},
\begin{align}\label{deno1}
    \sum_{n=1}^N \hat{m}(n)+\sum_{n=1}^N \hat{h}(n)\leq\frac{\alpha}{\lfloor k\rfloor}(\lceil r\rceil-\beta),
\end{align}
and the equality holds only if $\hat{\mu}(n)=\frac{\alpha}{\lfloor k\rfloor}$, $\forall n$, since the cases where $\beta\in\{0,1\}$ are already considered. $\hat{\mu}(n)=\frac{\alpha}{\lfloor k\rfloor}$, $\forall n$ specifies a set of homogeneous storage constraints which can be directly stored using the optimum storage mechanism provided in Section~\ref{improve} for homogeneous storage constraints.\footnote{The purpose of this step in the scheme is to divide the $\alpha$ fraction of all submodels that are $\lfloor k\rfloor$ coded into two MDS codes given by $(\lfloor k\rfloor,\lfloor r\rfloor)$ and $(\lfloor k\rfloor,\lceil r\rceil)$ such that all databases are filled and the coded parameters are replicated in the respective number of databases. This is satisfied by the scheme in Section~\ref{improve} for homogeneous storage constraints as explained next. $\hat{\mu}(n)$ corresponds to the storage allocated in database $n$ for all $\lfloor k\rfloor$ coded parameters. Based on the discussion in Section~\ref{improve}, any linear combination of $C_T(\lfloor k\rfloor,\lfloor r\rfloor)$ and $C_T(\lfloor k\rfloor,\lceil r\rceil)$ is achievable by storing $\beta$ and $1-\beta$ fractions of all $\lfloor k\rfloor$ coded parameters using $(\lfloor k\rfloor,\lfloor r\rfloor)$ and $(\lfloor k\rfloor,\lceil r\rceil)$ MDS codes, respectively, for any $\beta\in[0,1]$, given that $\lfloor k\rfloor\leq\lfloor r\rfloor-4$. Therefore, for the case where $\hat{\mu}(n)=\frac{\alpha}{\lfloor k\rfloor}$, $\forall n$, the $\alpha$ fraction of all submodels that are $\lfloor k\rfloor$ coded  in Table~\ref{codes} can be arbitrarily divided (arbitrary $\beta$) and encoded with the two MDS codes $(\lfloor k\rfloor,\lfloor r\rfloor)$ and $(\lfloor k\rfloor,\lceil r\rceil)$ to achieve a total cost of  $\alpha\beta C_T(\lfloor k\rfloor,\lfloor r\rfloor)+\alpha(1-\beta)C_T(\lfloor k\rfloor,\lceil r\rceil)$, while filling all databases.} Note that this case is not in the scope of Claim~2. For all other cases, \eqref{deno1} is satisfied with strict inequality, which proves $\mathcal{D}>0$. Therefore, it remains to prove that $\mathcal{N}\leq\mathcal{D}$ and $\mathcal{N}\geq0$, to prove $0\leq\hat{\gamma}\leq1$. Note that $\mathcal{N}\leq\mathcal{D}$ is equivalent to $\frac{\alpha\beta}{\lfloor k\rfloor}\lfloor r\rfloor\leq \frac{\alpha}{\lfloor k\rfloor}(\lceil r\rceil-\beta)-\sum_{n=1}^N \hat{h}(n)$. Let $V$ be defined as,
\begin{align}
    V=\sum_{n=1}^N \mathbf{1}_{\{\hat{\mu}(n)\geq \frac{\alpha\beta}{\lfloor k\rfloor}\}}.
\end{align}
Therefore, from the definition of $\hat{h}(n)$ in \eqref{hatmh} and since $\hat{\mu}(n)\leq\frac{\alpha}{\lfloor k\rfloor}$, $\forall n$ from \eqref{comb3},
\begin{align}
    \sum_{n=1}^N \hat{h}(n)\leq \frac{V\alpha}{\lfloor k\rfloor}-\frac{V\alpha\beta}{\lfloor k\rfloor}=\frac{V\alpha(1-\beta)}{\lfloor k\rfloor}.\label{in1no1}
\end{align}
Moreover, since the total available space in the $V$ databases must not exceed $\sum_{n=1}^N \hat{\mu}(n)=\frac{\alpha}{\lfloor k\rfloor}(\lceil r\rceil-\beta)$,
\begin{align}
    \sum_{n=1}^N \hat{h}(n)+\frac{V\alpha\beta}{\lfloor k\rfloor}\leq \frac{\alpha}{\lfloor k\rfloor}(\lceil r\rceil-\beta)\quad \implies\quad \sum_{n=1}^N \hat{h}(n)\leq \frac{\alpha}{\lfloor k\rfloor}(\lceil r\rceil-\beta)-\frac{V\alpha\beta}{\lfloor k\rfloor}\label{in2no1}
\end{align}
must be satisfied. The upper bound on $\sum_{n=1}^N \hat{h}(n)$ in \eqref{in1no1} is tighter than that of \eqref{in2no1} when $V\leq\lceil r\rceil-\beta$ and vice versa. Therefore, the highest upper bound on $\sum_{n=1}^N 
 \hat{h}(n)$ is given by,
\begin{align}
    \sum_{n=1}^N \hat{h}(n)\leq\begin{cases}
        \frac{\alpha(1-\beta)}{\lfloor k\rfloor}\lfloor r\rfloor, & \text{if $V\leq\lceil r\rceil-\beta$}\\
        \frac{\alpha}{\lfloor k\rfloor}(\lceil r\rceil-\beta)-\frac{\alpha\beta}{\lfloor k\rfloor}\lceil r\rceil, & \text{if $V>\lceil r\rceil-\beta$}
    \end{cases},
\end{align}
since $V\in\mathbb{Z}^+$, which gives the tightest upper bound on $\frac{\alpha}{\lfloor k\rfloor}(\lceil r\rceil-\beta)-\sum_{n=1}^N \hat{h}(n)$ for a general set of $\{\mu(n)\}_{n=1}^N$ as,
\begin{align}
    \frac{\alpha}{\lfloor k\rfloor}(\lceil r\rceil-\beta)-\sum_{n=1}^N \hat{h}(n)&\geq\begin{cases}
        \frac{\alpha}{\lfloor k\rfloor}(1-\beta+\beta\lfloor r\rfloor), & \text{if $V\leq \lceil r\rceil-\beta$}\\
        \frac{\alpha\beta}{\lfloor k\rfloor}\lceil r\rceil, & \text{if $V<\lceil r\rceil-\beta$}
    \end{cases}\\
    &\geq\frac{\alpha\beta}{\lfloor k\rfloor}\lfloor r\rfloor,
\end{align}
which proves $\mathcal{N}\leq\mathcal{D}$. To prove $\mathcal{N}\geq0$, we need to show that $\sum_{n=1}^N \hat{m}(n)\leq\frac{\alpha\beta}{\lfloor k\rfloor}\lfloor r\rfloor$. Let $Y$ be defined as,
\begin{align}
    Y=\sum_{n=1}^N \mathbf{1}_{\{\hat{\mu}(n)\geq \frac{\alpha(1-\beta)}{\lfloor k\rfloor}\}}.
\end{align}
Then, similar to \eqref{in1no1} and \eqref{in2no1}, we have
\begin{align}
    \sum_{n=1}^N \hat{m}(n)&\leq \frac{Y\alpha}{\lfloor k\rfloor}-\frac{Y\alpha(1-\beta)}{\lfloor k\rfloor}\label{in1mno1}\\
    \sum_{n=1}^N\hat{m}(n)&\leq \frac{\alpha}{\lfloor k\rfloor}(\lceil r\rceil-\beta)-\frac{Y\alpha(1-\beta)}{\lfloor k\rfloor}\label{in2mno1},
\end{align}
and \eqref{in1mno1} provides a tighter upper bound on $\sum_{n=1}^N \hat{m}(n)$ compared to \eqref{in2mno1} when $Y\leq\lceil r\rceil-\beta$, and vice versa. Therefore, the highest upper bound on $\sum_{n=1}^N \hat{m}(n)$ considering any arbitrary set of $\{\mu(n)\}_{n=1}^N$ is given by,
\begin{align}
    \sum_{n=1}^N \hat{m}(n)&\leq\begin{cases}
        \frac{\lfloor r\rfloor\alpha\beta}{\lfloor k\rfloor}, & \text{if $Y\leq\lceil r\rceil-\beta$}\\
        \frac{\alpha}{\lfloor k\rfloor}(\lceil r\rceil-\beta)-\frac{\lceil r\rceil\alpha(1-\beta)}{\lfloor k\rfloor}, & \text{if $Y>\lceil r\rceil-\beta$}
    \end{cases} \\
    &=\frac{\alpha\beta}{\lfloor k\rfloor}\lfloor r\rfloor
\end{align}
since $Y\in\mathbb{Z}^+$, proving $\mathcal{N}\geq0$, completing the proof of $0\leq\hat{\gamma}\leq1$. This proves Claim~2.
\end{Proof}

\textbf{Claim~3:}  For each $n\in\{1,\dotsc,N\}$, $\bar{\mu}_1(n)$ and $\bar{\mu}_2(n)$, in \eqref{a3} and \eqref{a4} (same as \eqref{mubar1} and \eqref{e2}) satisfy \eqref{eq3}-\eqref{eq4} and \eqref{neq3}-\eqref{neq4} for those $\alpha,\beta,\delta$ stated in Lemma~\ref{lem1}.

\begin{Proof}
    The proof od Claim~3 consists of the exact same steps as in the proof of Claim~2.
\end{Proof}

\section{Proof of Lemma~\ref{lem2}}

Here, we derive the optimum values of $\alpha,\beta,\delta$ that minimize \eqref{min} while satisfying \eqref{eq1}-\eqref{sum} for the case where $\alpha<1$. Note that the constraints in \eqref{alpha}-\eqref{delta} and \eqref{combb3} must be satisfied by $\alpha,\beta,\delta$ to guarantee \eqref{eq1}-\eqref{sum}. The following optimization problem is solved to obtain the optimum fractions:
\begin{align}
    \min \quad &C=\alpha\beta C_T(\lfloor k\rfloor,\lfloor r\rfloor)+\alpha(1-\beta) C_T(\lfloor k\rfloor,\lceil r\rceil)+(1-\alpha)\delta C_T(\lceil k\rceil,\lfloor r\rfloor)\nonumber\\
    &\qquad+(1-\alpha)(1-\delta) C_T(\lceil k\rceil,\lceil r\rceil)\\
    \text{s.t.} \quad &\frac{\lfloor k\rfloor}{k}(\lceil k\rceil-k)\leq\alpha<1\label{kk1}\\
    &\left[1-\frac{\lfloor k\rfloor}{k\alpha}(r-\lfloor r\rfloor)\right]^+\leq\beta\leq1\\
    &\left[1-\frac{\lceil k\rceil}{k(1-\alpha)}(r-\lfloor r\rfloor)\right]^+\leq\delta\leq1\\
    &\frac{\alpha}{\lfloor k\rfloor}(\lceil r\rceil-\beta)+\frac{1-\alpha}{\lceil k\rceil}(\lceil r\rceil-\delta)=p.\label{kk2}
\end{align}
where the total cost is explicitly given by,
\begin{align}
    \!C=\begin{cases}
        \frac{4\alpha\beta\lfloor r\rfloor}{\lfloor r\rfloor-\lfloor k\rfloor-1}+\frac{\alpha(1-\beta)(4\lceil r\rceil-2)}{\lceil r\rceil-\lfloor k\rfloor-2}+\frac{(1-\alpha)\delta(4\lfloor r\rfloor-2)}{\lfloor r\rfloor-\lceil k\rceil-2}+\frac{4(1-\alpha)(1-\delta)\lceil r\rceil}{\lceil r\rceil-\lceil k\rceil-1}, & \text{if $\lfloor r\rfloor-\lfloor k\rfloor$ is odd}\\
        \frac{\alpha\beta(4\lfloor r\rfloor-2)}{\lfloor r\rfloor-\lfloor k\rfloor-2}+\frac{4\alpha(1-\beta)\lceil r\rceil}{\lceil r\rceil-\lfloor k\rfloor-1}+\frac{4(1-\alpha)\delta\lfloor r\rfloor}{\lfloor r\rfloor-\lceil k\rceil-1}+\frac{(1-\alpha)(1-\delta)(4\lceil r\rceil-2)}{\lceil r\rceil-\lceil k\rceil-2}, & \text{if $\lfloor r\rfloor-\lfloor k\rfloor$ is even}
    \end{cases}.\label{expl}
\end{align}
We consider two cases for the two different total costs, and obtain the KKT conditions:

\textbf{Case~1: Odd $\lfloor r\rfloor-\lfloor k\rfloor$:} The Lagrangian function for this case is given by,\footnote{We treat the constraint $\alpha<1$ as $\alpha\leq1$ in the Lagrangian, and avoid the case $\alpha=1$ in the analysis.}
\begin{align}
    J&=C+\lambda_1(\alpha-1)+\lambda_2\left(\frac{\lfloor k\rfloor}{k}(\lceil k\rceil-k)-\alpha\right)+\lambda_3(\beta-1)+\lambda_4\left(1-\frac{\lfloor k\rfloor}{k\alpha}(r-\lfloor r\rfloor)-\beta\right)\nonumber\\
    &\quad-\lambda_5\beta+\lambda_6(\delta-1)+\lambda_7\left(1-\frac{\lceil k\rceil}{k(1-\alpha)}(r-\lfloor r\rfloor)-\delta\right)-\lambda_8\delta\nonumber\\
    &\quad+\lambda_9\left(\frac{\alpha}{\lfloor k\rfloor}(\lceil r\rceil-\beta)+\frac{1-\alpha}{\lceil k\rceil}(\lceil r\rceil-\delta)-p\right).\label{lag}
\end{align}
The KKT conditions for this case are,
\begin{align}
    &\frac{\partial J}{\partial\alpha}=\frac{4\beta\lfloor r\rfloor}{\lfloor r\rfloor-\lfloor k\rfloor-1}+\frac{(1-\beta)(4\lceil r\rceil-2)}{\lceil r\rceil-\lfloor k\rfloor-2}-\frac{\delta(4\lfloor r\rfloor-2)}{\lfloor r\rfloor-\lceil k\rceil-2}-\frac{4(1-\delta)\lceil r\rceil}{\lceil r\rceil-\lceil k\rceil-1}+\lambda_1-\lambda_2\nonumber\\
    &\quad \qquad+\lambda_4\frac{\lfloor k\rfloor}{k\alpha^2}(r-\lfloor r\rfloor)-\lambda_7\frac{\lceil k\rceil}{k(1-\alpha)^2}(r-\lfloor r\rfloor)+\lambda_9\left(\frac{\lceil r\rceil-\beta}{\lfloor k\rfloor}-\frac{\lceil r\rceil-\delta}{\lceil k\rceil}\right)=0\\
    &\frac{\partial J}{\partial\beta}=\frac{4\alpha\lfloor r\rfloor}{\lfloor r\rfloor-\lfloor k\rfloor-1}-\frac{\alpha(4\lceil r\rceil-2)}{\lceil r\rceil-\lfloor k\rfloor-2}+\lambda_3-\lambda_4-\lambda_5-\lambda_9\frac{\alpha}{\lfloor k\rfloor}=0\\
    &\frac{\partial J}{\partial\delta}=\frac{(1-\alpha)(4\lfloor r\rfloor-2)}{\lfloor r\rfloor-\lceil k\rceil-2}-\frac{4(1-\alpha)\lceil r\rceil}{\lceil r\rceil-\lceil k\rceil-1}+\lambda_6-\lambda_7-\lambda_8-\lambda_9\frac{(1-\alpha)}{\lceil k\rceil}=0\\
    &\lambda_1(\alpha-1)=0, \quad \lambda_2\left(\frac{\lfloor k\rfloor}{k}(\lceil k\rceil-k)-\alpha\right)=0,\quad \lambda_3(\beta-1)=0,\\
    &\lambda_4\left(1-\frac{\lfloor k\rfloor}{k\alpha}(r-\lfloor r\rfloor)-\beta\right)=0,\quad \lambda_5\beta=0,\quad\lambda_6(\delta-1)=0,\\
    &\lambda_7\left(1-\frac{\lceil k\rceil}{k(1-\alpha)}(r-\lfloor r\rfloor)-\delta\right)=0,\quad\lambda_8\delta=0\\
    &\lambda_i\geq0,\quad i\in\{1,\dotsc,8\},
\end{align}
and \eqref{kk1}-\eqref{kk2}. Solving the above KKT conditions results in the optimum $\alpha,\beta,\delta$ stated in \eqref{odd1}-\eqref{odd2} in Theorem~\ref{thm1}, along with the minimum total cost.

\textbf{Case~2: Even $\lfloor r \rfloor-\lfloor k\rfloor$:} The Lagrangian function for this case is the same as \eqref{lag}, with the respective total cost $C$ from \eqref{expl}. The KKT conditions for this case are,
\begin{align}
    &\frac{\partial J}{\partial\alpha}=\frac{\beta(4\lfloor r\rfloor-2)}{\lfloor r\rfloor-\lfloor k\rfloor-2}+\frac{4(1-\beta)\lceil r\rceil}{\lceil r\rceil-\lfloor k\rfloor-1}-\frac{4\delta\lfloor r\rfloor}{\lfloor r\rfloor-\lceil k\rceil-1}-\frac{(1-\delta)(4\lceil r\rceil-2)}{\lceil r\rceil-\lceil k\rceil-2}+\lambda_1-\lambda_2\nonumber\\
    &\quad \qquad+\lambda_4\frac{\lfloor k\rfloor}{k\alpha^2}(r-\lfloor r\rfloor)-\lambda_7\frac{\lceil k\rceil}{k(1-\alpha)^2}(r-\lfloor r\rfloor)+\lambda_9\left(\frac{\lceil r\rceil-\beta}{\lfloor k\rfloor}-\frac{\lceil r\rceil-\delta}{\lceil k\rceil}\right)=0\\
    &\frac{\partial J}{\partial\beta}=\frac{\alpha(4\lfloor r\rfloor-2)}{\lfloor r\rfloor-\lfloor k\rfloor-2}-\frac{4\alpha\lceil r\rceil}{\lceil r\rceil-\lfloor k\rfloor-1}+\lambda_3-\lambda_4-\lambda_5-\lambda_9\frac{\alpha}{\lfloor k\rfloor}=0\\
    &\frac{\partial J}{\partial\delta}=\frac{4(1-\alpha)\lfloor r\rfloor}{\lfloor r\rfloor-\lceil k\rceil-1}-\frac{(1-\alpha)(4\lceil r\rceil-2)}{\lceil r\rceil-\lceil k\rceil-2}+\lambda_6-\lambda_7-\lambda_8-\lambda_9\frac{(1-\alpha)}{\lceil k\rceil}=0\\
    &\lambda_1(\alpha-1)=0, \quad \lambda_2\left(\frac{\lfloor k\rfloor}{k}(\lceil k\rceil-k)-\alpha\right)=0,\quad \lambda_3(\beta-1)=0,\\
    &\lambda_4\left(1-\frac{\lfloor k\rfloor}{k\alpha}(r-\lfloor r\rfloor)-\beta\right)=0,\quad \lambda_5\beta=0,\quad\lambda_6(\delta-1)=0,\\
    &\lambda_7\left(1-\frac{\lceil k\rceil}{k(1-\alpha)}(r-\lfloor r\rfloor)-\delta\right)=0,\quad\lambda_8\delta=0\\
    &\lambda_i\geq0,\quad i\in\{1,\dotsc,8\},
\end{align}
and \eqref{kk1}-\eqref{kk2}. Solving the above KKT conditions results in the optimum $\alpha,\beta,\delta$ stated in \eqref{even1}-\eqref{even2} in Theorem~\ref{thm1}.

\bibliographystyle{unsrt}
\bibliography{references-journal}

\end{document}